\documentclass[preprintnumbers,superscriptaddress,showkeys,showpacs,byrevtex]{revtex4}
\usepackage{amsmath,amsfonts,amssymb,amscd,amsxtra,amsthm}
\usepackage{graphicx}
\usepackage{bm}
\usepackage{color}
\usepackage{graphics}
\usepackage{amsmath}
\usepackage{epsfig}
\usepackage{epsf}
\usepackage{epstopdf}
\usepackage{multirow}
\begin{document}
\preprint{CYCU-HEP-15-10}
\preprint{PKNU-NuHaTh-2015-03}
\title{Consistency check of charged hadron multiplicities and fragmentation functions in SIDIS}
\author{Dong-Jing Yang}
\email[E-mail: ]{djyang@std.ntnu.edu.tw}
\affiliation{Department of Physics, National Taiwan Normal University, Taipei 11677, Taiwan}
\author{Fu-Jiun Jiang}
\email[E-mail: ]{fjjiang@ntnu.edu.tw}
\affiliation{Department of Physics, National Taiwan Normal University, Taipei 11677, Taiwan}
\author{Wen-Chen Chang}
\email[E-mail: ]{changwc@phys.sinica.edu.tw}
\affiliation{Institute of Physics, Academia Sinica, Taipei 11529, Taiwan}
\author{Chung-Wen Kao}
\email[E-mail  (Corresponding Author): ]{cwkao@cycu.edu.tw}
\affiliation{Department of Physics and Center for High Energy Physics, Chung-Yuan Christian University, \\Chung-Li 32023, Taiwan}
\author{Seung-il Nam}
\email[E-mail: ]{sinam@pknu.ac.kr}
\affiliation{Department of Physics, Pukyong National University (PKNU), Busan 608-737, Republic of Korea}
\affiliation{Asia Pacific Center for Theoretical Physics (APCTP), Pohang 790-784,
Republic of Korea}
\date{\today}
\begin{abstract}
We derived the conditions on certain combinations of integrals
of the fragmentation functions of pion using HERMES data of the sum for the charged pion multiplicities from semi-inclusive
deep-inelastic scattering (SIDIS) off the deuteron target. In our derivation the nucleon parton distribution functions (PDFs) are
assumed to be isospin SU(2) symmetric. Similar conditions have also been obtained for the fragmentation functions (FFs) of kaon by
the sum of charged kaon multiplicities as well.
We have chosen several FFs to study the impact of those conditions we have derived. Among those FFs, only that produced in the nonlocal chiral-quark model (NL$\chi$QM) constantly satisfy the conditions.
Furthermore, the ratios of the strange PDFs$S(x)$ and the nonstrange PDFs $Q(x,Q^2)$  extracted from the
charged pion and kaon multiplicities differ from each other significantly.
Finally, we demonstrate that the HERMES pion multiplicity data is unlikely to be compatible with the two widely-used
PDFs, namely CTEQ6M and NNPDF3.0.
\end{abstract}
\maketitle

\section{Introduction}
Semi-inclusive deep-inelastic scattering, $l+N\to l+h+X$ (SIDIS) plays an important role in the study of the fragmentation functions (FFs).
In particular it provides valuable information about the flavour dependence of fragmentation functions which cannot be obtained from
$e^{+}e^{-}$ annihilation data. According to the leading order (LO) QCD calculation, the sum of the charged pion ($\pi$) multiplicities of
SIDIS off a deuteron ($D$) target, which is denoted by $M_{D}^{\pi}(x,Q^2)$, is given by
\begin{eqnarray}
M^{\pi}_{D}(x,Q^2)&\equiv&M^{\pi^{+}}_{D}(x,Q^2)+M^{\pi^{-}}_{D}(x,Q^2)=\frac{dN^{\pi}(x,Q^2)}{dN^\mathrm{DIS}(x,Q^2)}
\cr
&=&\frac{\sum_{q}e_{q}^2\left[q^{p}(x,Q^2)+\bar{q}^{p}(x,Q^2)+q^{n}(x,Q^2)+\bar{q}^{n}(x,Q^2)\right]\int^{z_\mathrm{max}}_{z_\mathrm{min}}D_{q}^{\pi}(z,Q^2)dz}
{\sum_{q}e_{q}^2\left[q^{p}(x,Q^2)+\bar{q}^{p}(x,Q^2)+q^{n}(x,Q^2)+\bar{q}^{n}(x,Q^2)\right]}.
\label{Eq:ori2}
\end{eqnarray}
Here $q=(u,d,s)$ and $e_q$ are the considered quarks and the corresponding charges, respectively.
In addition, $q^{i}(x,Q^2)$ with $i \in \{p,n\}$ are the relevant nucleon parton distribution functions (PDFs) with momentum fraction $x$ and
momentum transferred squared $Q^2$. Notice the superscripts $p$ and $n$ denote proton and neutron.
The $z$ is the momentum fraction of the initial quark in the fragmented hadron. $z_\mathrm{max}$ and $z_\mathrm{min}$ are usually set by the
experimental acceptance. Finally $D_{q}^{\pi}$ in Eq.~(\ref{Eq:ori2}) is defined in terms of FFs as well and takes the the
following form
\begin{equation}
D_{q}^{\pi}(z,Q^2)=D_{q}^{\pi^{+}}(z,Q^2)+D_{q}^{\pi^{-}}(z,Q^2).
\end{equation}
In the derivation of Eq. (\ref{Eq:ori2}) we have applied the relations
\begin{equation}
D^{\pi^{+}}_{q}(z,Q^2) =D^{\pi^{-}}_{\bar{q}}(z,Q^2),\,\,\,\,\,D^{\pi^{-}}_{q}(z,Q^2) =D^{\pi^{+}}_{\bar{q}}(z,Q^2),
\end{equation}
since $\pi^{+}\rightarrow \pi^{-}$ and $q\rightarrow \bar{q}$ under the charge conjugation.

Furthermore, if we assume that the parton distribution functions (PDFs) are exactly SU(2) isospin symmetric, we have
\begin{eqnarray}
u^{p}(x,Q^2)&=&d^{n}(x,Q^2),\,\,\,\,d^{p}(x,Q^2)=u^{n}(x,Q^2),
\,\,\,\,s^{p}(x,Q^2)=s^{n}(x,Q^2),\cr
\bar{u}^{p}(x,Q^2)&=&\bar{d}^{n}(x,Q^2),\,\,\,\,\bar{d}^{p}(x,Q^2)=\bar{u}^{n}(x,Q^2),\,\,\,\,\bar{s}^{p}(x,Q^2)=s^{n}(x,Q^2).
\label{Eq:CSB}
\end{eqnarray}
The use of Eq.~(\ref{Eq:CSB}) to replace the PDFs in Eq.~(\ref{Eq:ori2}) will  lead to the following formula
\begin{equation}
M^{\pi}_{D}(x,Q^2)=M^{\pi^{+}}_{D}(x,Q^2)+M^{\pi^{-}}_{D}(x,Q^2)=\frac{dN^{\pi}(x,Q^2)}{dN^\mathrm{DIS}(x,Q^2)}
=\frac{Q(x,Q^2)D^{\pi}_{Q}(Q^2)+S(x,Q^2)D^{\pi}_{S}(Q^2)}{5Q(x,Q^2)+2S(x,Q^2)}.
\label{Eq:pion}
\end{equation}
Here, we define $S(x,Q^2)=s^{p}(x,Q^2)+\bar{s}^{p}(x,Q^2)$ and $Q(x)=u^{p}(x,Q^2)+d^{p}(x,Q^2)+\bar{u}^{p}(x,Q^2)+\bar{d}^{p}(x,Q^2)$.
$D^{\pi}_{Q}(Q^2)$ and $D^{\pi}_{S}(Q^2)$ are also given as follows:
\begin{eqnarray}
D^{\pi}_{Q}(Q^2)&=&4\int_{z_\mathrm{min}}^{z_\mathrm{max}}D^{\pi^{+}}_{u}(z,Q^2)dz+\int_{z_\mathrm{min}}^{z_\mathrm{max}}D^{\pi^{+}}_{d}(z,Q^2)dz
+4\int_{z_\mathrm{min}}^{z_\mathrm{max}}D^{\pi^{-}}_{u}(z,Q^2)dz+\int_{z_\mathrm{min}}^{z_\mathrm{max}}D^{\pi^{-}}_{d}(z,Q^2)dz \nonumber \\
&=&4D^{\pi^{+}}_{u}(Q^2)+D^{\pi^{+}}_{d}(Q^2)+4D^{\pi^{-}}_{u}(Q^2)+D^{\pi^{-}}_{d}(Q^2),\cr
D^{\pi}_{S}(Q^2)&=&2\int_{z_\mathrm{min}}^{z_\mathrm{max}}D^{\pi^{+}}_{s}(z,Q^2)dz
+2\int_{z_\mathrm{min}}^{z_\mathrm{max}}D^{\pi^{-}}_{s}(z,Q^2)dz=2D^{\pi^{+}}_{s}(Q^2)+2D^{\pi^{-}}_{s}(Q^2).
\label{Eq:ABPi}
\end{eqnarray}
Note that the FFs of $u\rightarrow \pi^{+}$ and $d\rightarrow \pi^{-}$ are favored ones, indicating that the quark fragments into
the hadron, whose constituent content has the same flavour. The favored FFs are supposed to be much larger than the unfavored ones.
Hence, one concludes that the magnitudes of $D^{\pi}_{Q}(Q^2)$
should be larger than those of $D^{\pi}_{S}(Q^2)$
at the same $Q^2$ value.

A similar relation for the kaon multiplicity of SIDIS off deuteron can be derived as follows:
\begin{equation}
M^{K}_{D}(x,Q^2)\equiv M^{K^{+}}_{D}(x,Q^2)+M^{K^{-}}_{D}(x,Q^2)=\frac{dN^{K}(x,Q^2)}{dN^\mathrm{SIDIS}(x,Q^2)}
=\frac{Q(x,Q^2)D^{K}_{Q}(Q^2)+S(x,Q^2)D^{K}_{S}(Q^2)}{5Q(x,Q^2)+2S(x,Q^2)},
\label{Eq:kaon}
\end{equation}
where $D^{K}_{Q}(Q^2)$ and $D^{K}_{S}(Q^2)$ are defined by
\begin{eqnarray}
D^{K}_{Q}(Q^2)&=4&\int_{z_\mathrm{min}}^{z_\mathrm{max}}D^{K^{+}}_{u}(z,Q^2)dz+\int_{z_\mathrm{min}}^{z_\mathrm{max}}D^{K^{+}}_{d}(z,Q^2)dz
+4\int_{z_\mathrm{min}}^{z_\mathrm{max}}D^{K^{-}}_{u}(z,Q^2)dz+\int_{z_\mathrm{min}}^{z_\mathrm{max}}D^{K^{-}}_{d}(z,Q^2)dz. \nonumber \\
&=&4D^{K^{+}}_{u}(Q^2)+D^{K^{+}}_{d}(Q^2)+4D^{K^{-}}_{u}(Q^2)+D^{K^{-}}_{d}(Q^2),
\cr
D^{K}_{S}(Q^2)&=2&\int_{z_\mathrm{min}}^{z_\mathrm{max}}D^{K^{+}}_{s}(z,Q^2)dz
+2\int_{z_\mathrm{min}}^{z_\mathrm{max}}D^{K^{-}}_{s}(z,Q^2)dz
\cr
&=&2D^{K^{+}}_{s}(Q^2)+2D^{K^{-}}_{s}(Q^2).
\label{Eq:AB_K}
\end{eqnarray}
Among the FFs appearing in Eq.~(\ref{Eq:AB_K}), only those of $u\to K^{+}$ and $s\to K^{-}$ are favored ones.
Therefore, unlike the pion
case, $D^{K}_{Q}(Q^2)$ is not necessarily larger than $D^{K}_{S}(Q^2)$. There are attempts to extract the strange-quark PDF
from the data of $M^{K}_{D}(x,Q^2)$ by using Eq.~(\ref{Eq:kaon}), but not without controversy~\cite{strange1,strange2,strange3}.
We will comment on this issue later.

In this article we analyze the data of HERMES pion and kaon multiplicities~\cite{pion,kaon2}
according to Eqs.~(\ref{Eq:pion}) and (\ref{Eq:kaon}). In Section II we use Eq.~(\ref{Eq:pion}) to derive the
conditions of constraints on
$D^{\pi}_{Q}(Q^2)$ and $D^{\pi}_{S}(Q^2)$. Furthermore, in the same section we examine whether these derived constraints are
satisfied by the FFs resulting from several parametrizations and models.
We repeat the same analysis for the kaon case by using Eq.~(\ref{Eq:kaon}) in section III.
In Section IV we discuss the inconsistency between $S(x,Q^2)/Q(x,Q^2)$ extracted from the pion multiplicities and kaon multiplicities.
After section IV, we choose certain parametrizations of PDFs to determine the corresponding values of
$D^{\pi}_{Q}$ and $D^{\pi}_{S}$, assuming that they are not sensitive to $Q^2$. Finally we summarize our results and make our conclusions in Section VI.

\section{The conditions on the fragmentation functions of charged pions}
\begin{figure}[t]
\begin{tabular}{ccc}
\includegraphics[width=5.2cm]{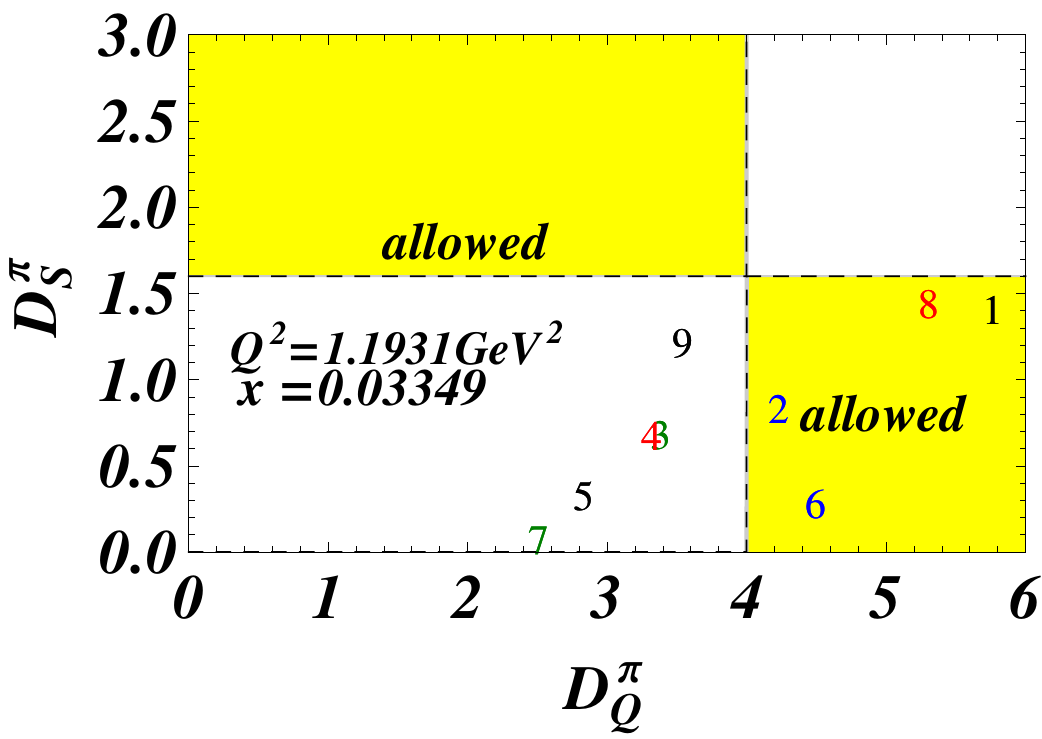}
\includegraphics[width=5.2cm]{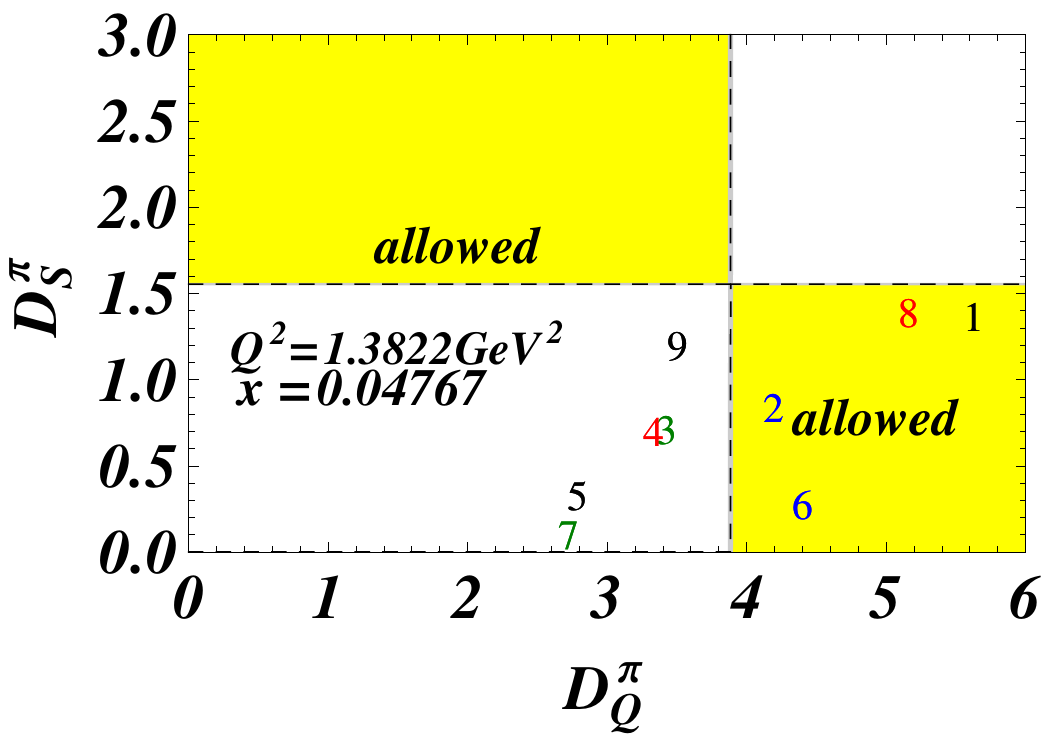}
\includegraphics[width=5.2cm]{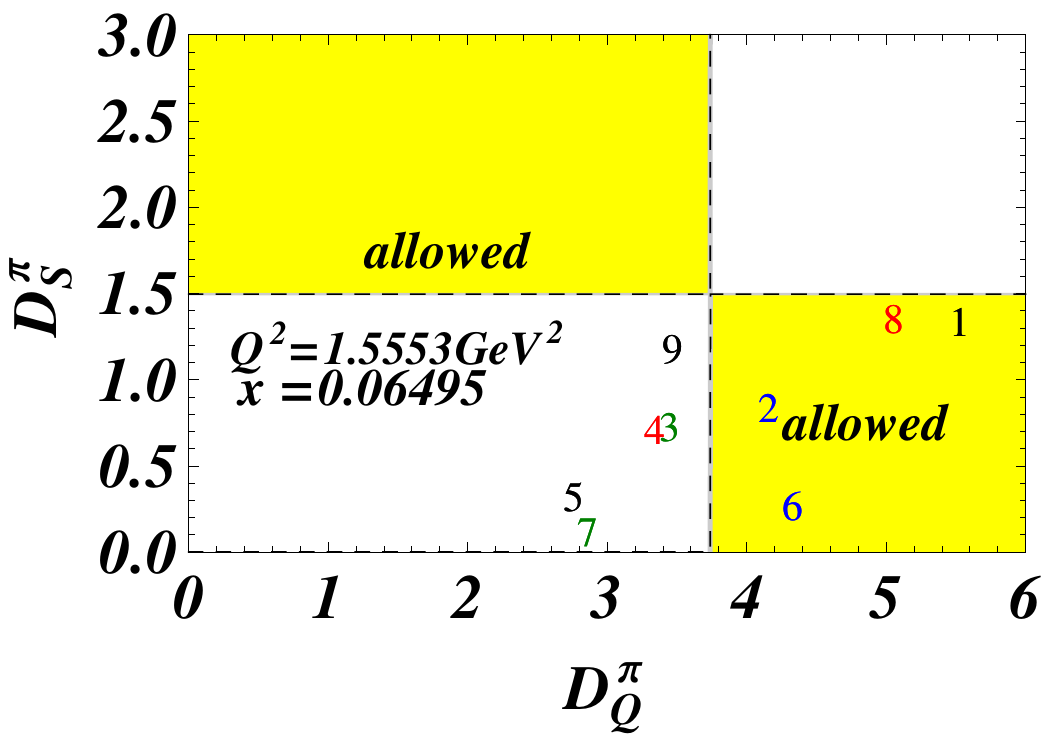}
\end{tabular}
\begin{tabular}{ccc}
\includegraphics[width=5.2cm]{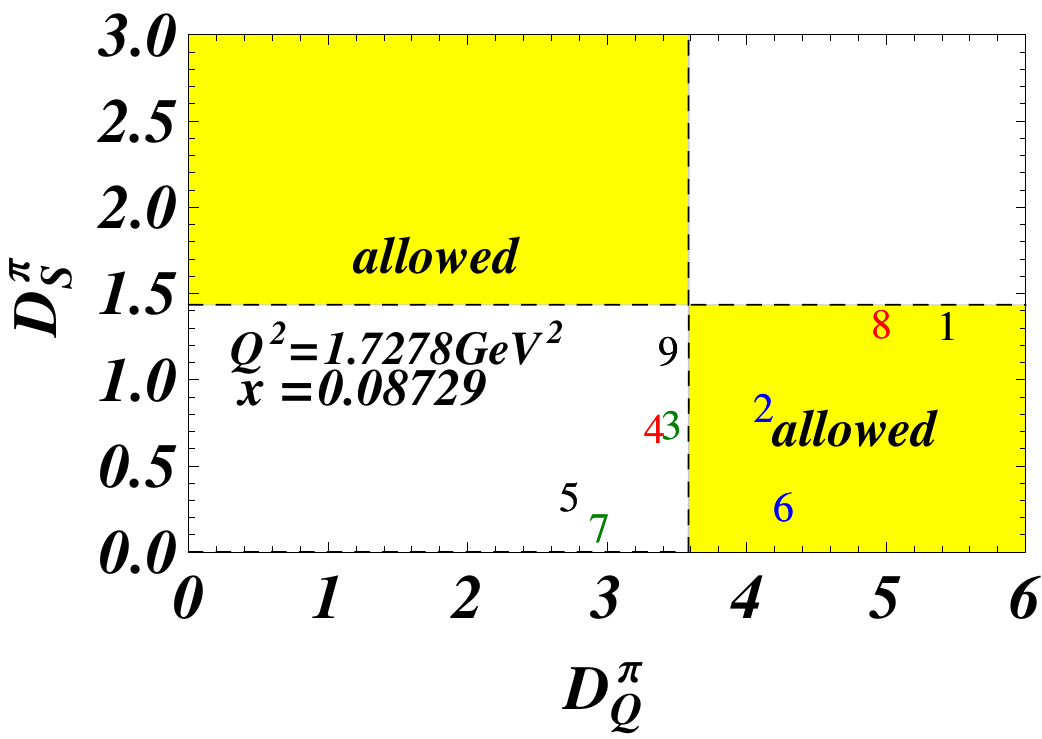}
\includegraphics[width=5.2cm]{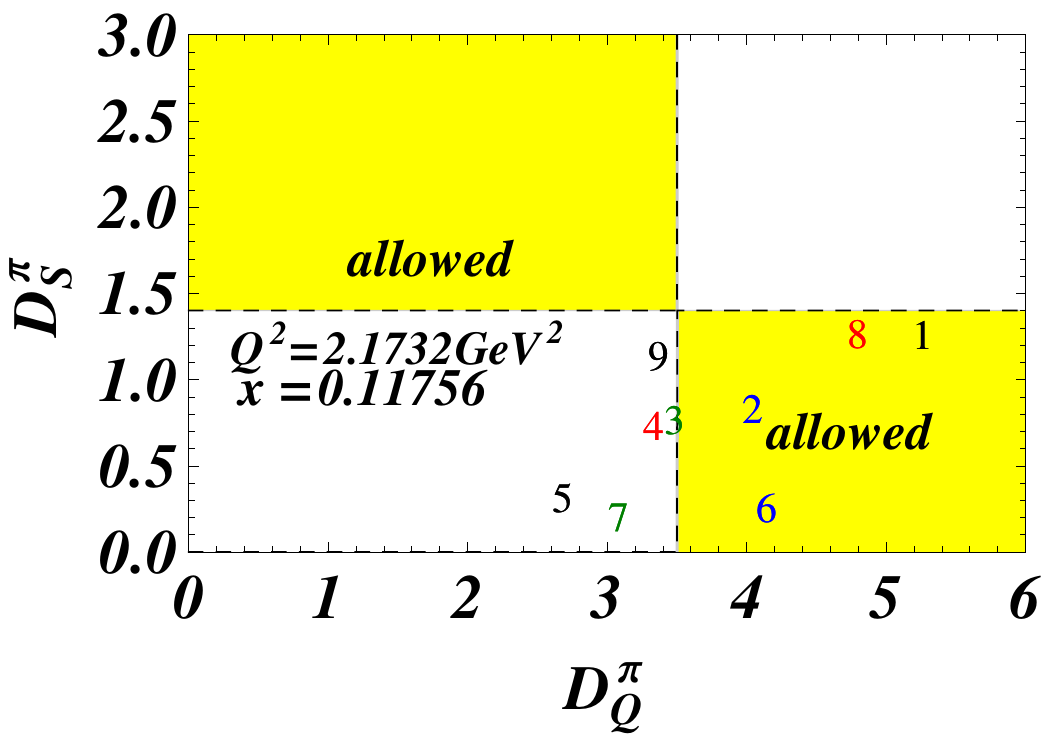}
\includegraphics[width=5.2cm]{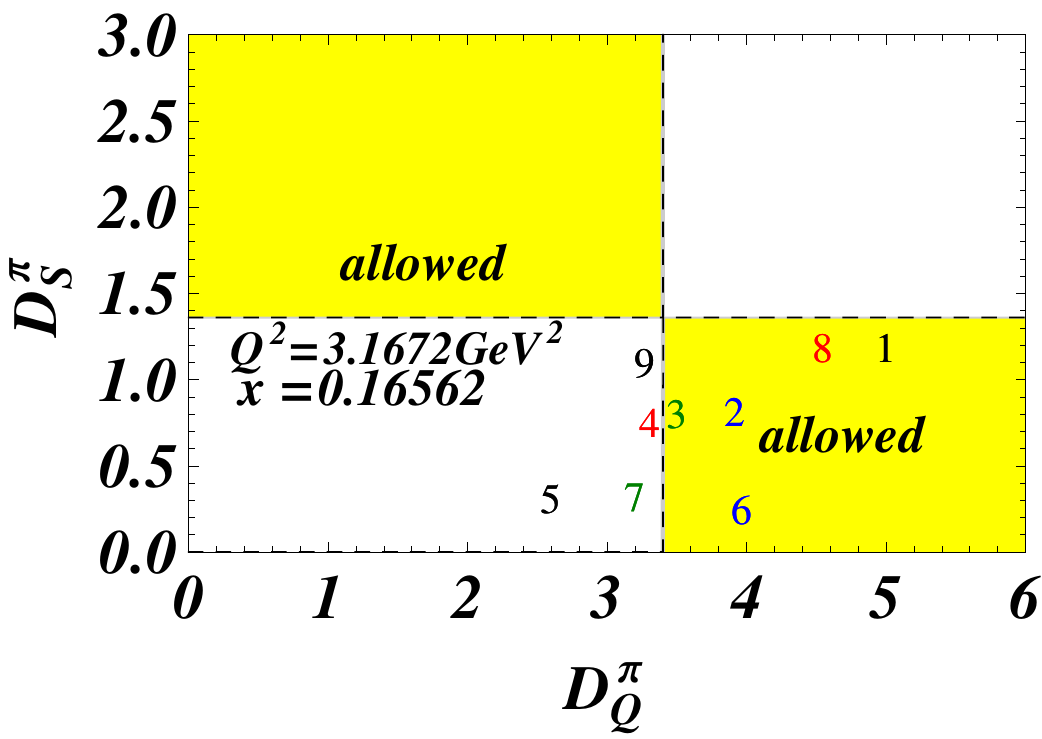}
\end{tabular}
\begin{tabular}{ccc}
\includegraphics[width=5.2cm]{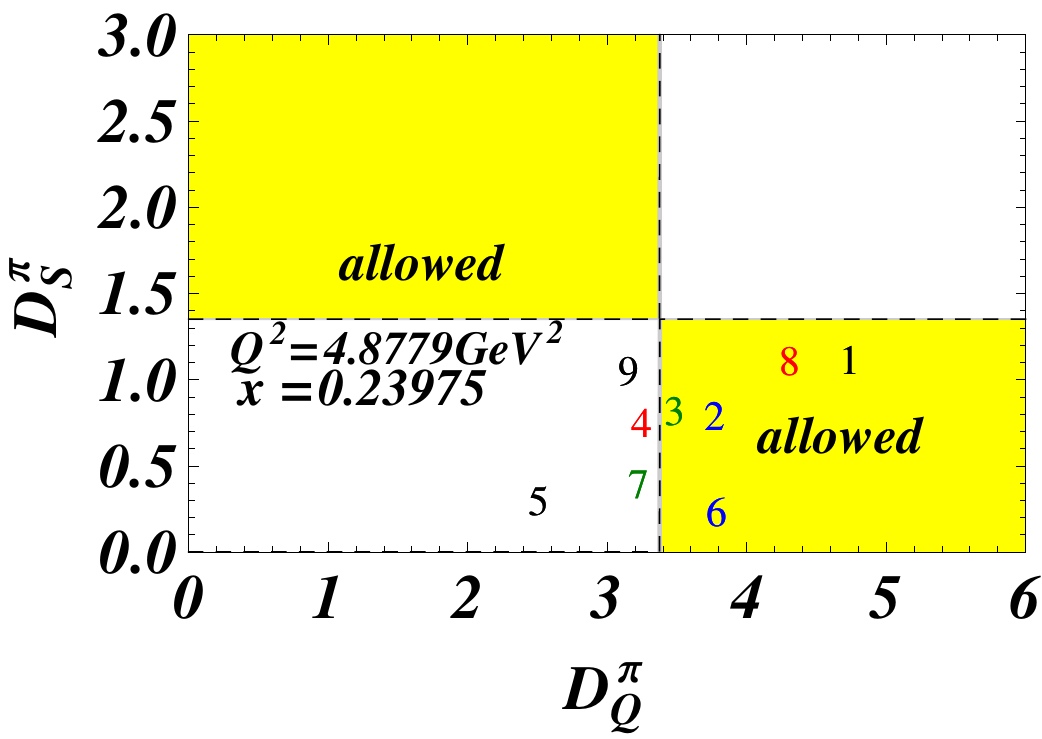}
\includegraphics[width=5.2cm]{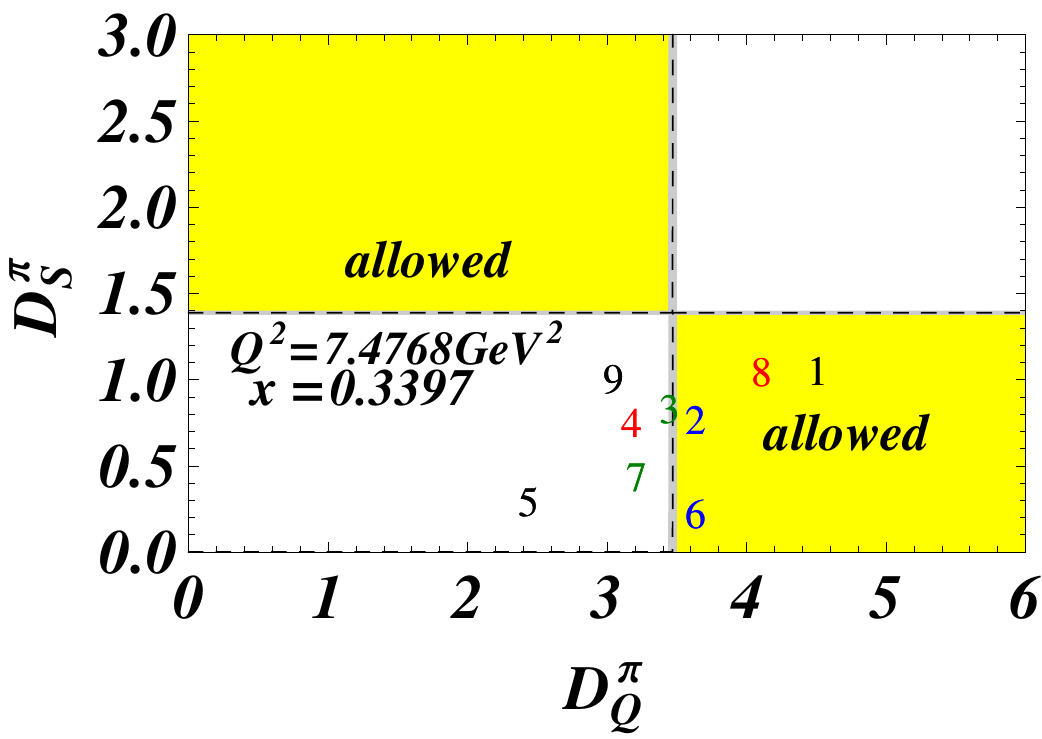}
\includegraphics[width=5.2cm]{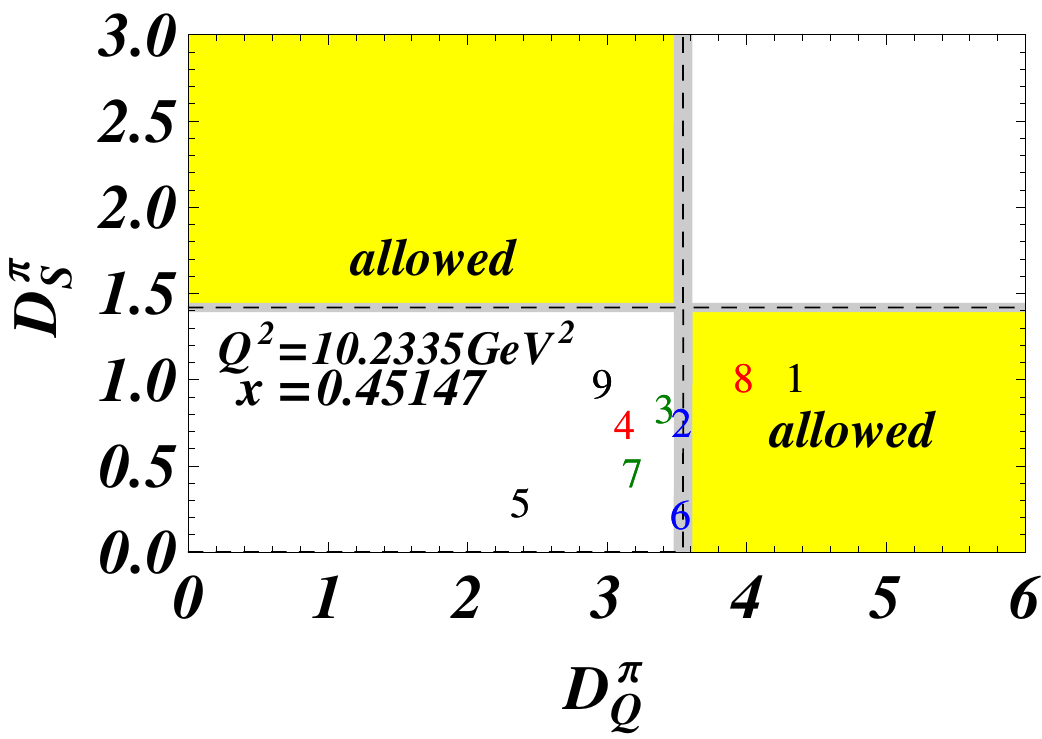}
\end{tabular}
\caption{The values of $D^{\pi}_{Q}(Q^2)$ and $D^{\pi}_{S}(Q^2)$ defined in Eq.~(\ref{Eq:ABPi}) from various fragmentation functions: HKNS parametrization at LO (1), HKNS parametrization at NLO (2), DSS parametrization at LO (3), DSS parametrization at NLO (4),  NJL-Jet model (5), nonlocal chiral-quark model (6),  AKK08 parametrization (7), SKMA parametrization (8), and DSEHPS parametrization (9). The yellow blocks represent the allowed regions experimentally. The grey bands stand for the areas corresponding to the estimated uncertainties
of $M_D^{\pi}$.}
\label{Fig:pion}
\end{figure}
In this section we will derive the condition on the integrals of the FFs, $D^{\pi}_{Q}(Q^2)$ and $D^{\pi}_{S}(Q^2)$. Let us rewrite Eq.~(\ref{Eq:pion}) into the following form:
\begin{equation}
S(x,Q^2)=\left[\frac{5M^{\pi}_{D}(x,Q^2)-D^{\pi}_{Q}(Q^2)}{D^{\pi}_{S}(Q^2)-2M^{\pi}_{D}(x,Q^2)}\right]Q(x,Q^2).
\label{Eq:S/Q_pion}
\end{equation}
It is obvious that the following relations must be true since both of $S(x,Q^2)$ and $Q(x,Q^2)$ must be positive,
\begin{equation}
\frac{D^{\pi}_{S}(Q^2)}{2}<M^{\pi}_{D}(x,Q^2)\le\frac{D^{\pi}_{Q}(Q^2)}{5}
\,\,\,\,\mathrm{or}\,\,\,\,
\frac{D^{\pi}_{Q}(Q^2)}{5}\le M^{\pi}_{D}(x,Q^2)<\frac{D^{\pi}_{S}(Q^2)}{2}.
\label{constaint1}
\end{equation}
In other words, $D^{\pi}_{Q}(Q^2)$ and $D^{\pi}_{S}(Q^2)$ must satisfy the following relations:
\begin{equation}
D^{\pi}_{Q}(Q^2)\ge 5M^{\pi}_{D}(x,Q^2),\,\,\,D^{\pi}_{S}(Q^2)<2M^{\pi}_{D}(x,Q^2)
\,\,\,\,\mathrm{or}\,\,\,\,
D^{\pi}_{Q}(Q^2)\le 5M^{\pi}_{D}(x,Q^2),\,\,\,D^{\pi}_{S}(Q^2)>2M^{\pi}_{D}(x,Q^2).
\label{constraint2}
\end{equation}
Note that these conditions are independent of the explicit forms of PDFs, $S(x,Q^2)$ and $Q(x,Q^2)$.

\begin{table}[t]
\begin{tabular}{|c|c|c|c|c|}
\hline
Data &$x$ & $Q^2$ & $M^{\pi}_{D}$ &  $M^{K}_{D}$ \\ \hline
A&0.03349 &1.1931 &0.800 &0.132  \\  \hline
B&0.04767 &1.3822 &0.777 &0.126 \\ \hline
C&0.06495 &1.5553 &0.748 &0.121  \\  \hline
D&0.08729 &1.7278 &0.717 &0.107 \\ \hline
E&0.11756 &2.1732 &0.701 &0.106 \\ \hline
F&0.16562 &3.1672 &0.680 &0.104 \\ \hline
G&0.23975 &4.8779 &0.676 &0.103 \\ \hline
H&0.3397 &7.4768 &0.694 &0.111 \\ \hline
J&0.45147 &10.2355 &0.709 &0.113  \\
\hline
\end{tabular}
\caption{HERMES data of the pion and kaon multiplicities off the deuteron target ($M_D^{\pi,K}$) at different $x$ and $Q^2$ values~\cite{pion}.}
\label{Table:HERMES}
\end{table}
\begin{table}[b]
\begin{tabular}{|c|c|c|c|c|c|}
\hline
Number& Name & Order  & Category &  Reference \\ \hline
1&HKNS &LO &Parametrization & \cite{HKNS} \\  \hline
2&HKNS &NLO &Parametrization  &    \cite{HKNS} \\ \hline
3&DSS &LO &Parametrization  &  \cite{DSS} \\  \hline
4&DSS &NLO &Parametrization  &  \cite{DSS}\\ \hline
5&NJL-Jet &$-$&Model & \cite{Matevosyan:2010hh,Matevosyan:2011ey} \\ \hline
6&NL$\chi$QM &$-$ &Model & \cite{Nam:2012af,Nam:2011hg,Yang:2013cza}\\ \hline
7&AKK08 & NLO &Parametrization& \cite{AKK} \\ \hline
8&SKMA & LO &Parametrization &   \cite{SKMA}\\ \hline
9&DSEHS & NLO &Parametrization   & \cite{DSEHS}  \\
\hline
\end{tabular}
\caption{Various fragmentation functions chosen in this article. See the text for details.}
\label{Table:FF}
\end{table}
The experimental results for the charged pion multiplicity $M^{\pi}_{D}$ reported by HERMES~\cite{pion} are listed in
Table~\ref{Table:HERMES}. 
In particular, we have assigned an alphbet character to each data point.
Furthermore, in Table~\ref{Table:HERMES} the related integral limits are $z_\mathrm{min}=0.2$ and $z_\mathrm{max}=0.8$.
The parametrization and models of FFs chosen in our study are listed in Table~\ref{Table:FF}.
Each FF of the considered parametrization or model is assigned by a number $(1\sim9)$ for convenience.
We have determined the corresponding $D^{\pi}_{Q}(Q^2)$ and $D^{\pi}_{S}(Q^2)$, and the results are listed in
Table~\ref{Table:pion}.
Since our analysis is basesd on the LO QCD formula, in principle one should choose only the LO
parametrizations of FFs.
Nevertheless we still consider several NLO parametrizations for comparison. \\

Notice each data point in Table~\ref{Table:HERMES} is taken at different $x$ and $Q^2$ values.
Hence to examine whether the condition Eq.~(\ref{constraint2}) is satisfied or not, we have carried out the analysis
at every data point separately. The numerical results are presented in
Fig.~\ref{Fig:pion}. The yellow block at r.h.s represents the allowed region satisfying
$D^{\pi}_{Q}(Q^2)\ge 5M^{\pi}_{D}(x,Q^2)$ and $D^{\pi}_{S}(Q^2)<2M^{\pi}_{D}(x,Q^2)$, whereas the one at l.h.s. denotes the allowed region
for $D^{\pi}_{Q}(Q^2)\le 5M^{\pi}_{D}(x,Q^2)$ and $D^{\pi}_{S}(Q^2)>2M^{\pi}_{D}(x,Q^2)$. In principle the values of $D^{\pi}_{Q}(Q^2)$
and $D^{\pi}_{S}(Q^2)$ should be within either block.
However, as can been seen in Fig.~\ref{Fig:pion}, our
results show a surprise.

We observe that only the Hirai--Kumano--Nagai--Sudoh (HKNS) parametrization for LO ($1$) and NLO ($2$)~\cite{HKNS},
LO Soleymaninia--Khorramian--Moosavinejad--Arbabifar (SKMA) parametrization~\cite{SKMA},
and NL$\chi$QM  ($6$) satisfy Eq.~(\ref{constraint2}) at every data point. They are all located inside the r.h.s. block.
Actually at data point (J), the values of the NLO de Florian-Sassotand-Stratmann (DSS) parametrization ($4$)~\cite{DSS}
and NL$\chi$QM  ($6$) both locate at the left brink of the r.h.s. block. It means that their values of $D^{\pi}_{Q}(Q^2)$
are very close to $5M^{\pi}_{D}$, as a result the corresponding $S(x,Q^2)/Q(x,Q^2)$ become small according
to Eq.~(\ref{Eq:S/Q_pion}).
On the other hand, the values of the LO HKNS parametrization ($1$) and LO SKMA parametrization
at low-$x$ region are near the top edge of the r.h.s. block, indicating that their values of $D^{\pi}_{S}(Q^2)$ are very
close to $2M^{\pi}_{D}(x,Q^2)$. Therefore, the corresponding $S(x,Q^2)/Q(x,Q^2)$ become huge. We also notice that the values of
$D^{\pi}_{Q}(Q^2)$ from LO HKNS parametrization ($1$) are particularly large. Moreover, the DSS parametrization values at low $Q^2$
are out of the allowed region, but they become more close to the left edge of the r.h.s. block as $Q^2$ and $x$ increase.
Besides, we notice that NJL-Jet model ($5$),
the de Florian-Sassotand--Epele--Hern\'anderz-Pinto--Stratmann (DSEHS, 9) and Albino--Kniehl--Kramer 08 (AKK08, 7)~\cite{AKK}
parametrizations fail to meet the requirement of Eq.~ (\ref{constraint2}) at all data points. Hence the corresponding
values of $S(x,Q^2)/Q(x,Q^2)$ are always negative. Here we want to make a remark on DSEHS parametrization in particular. DSEHS
parametrization is the most updated result of global analysis based on various experimental data including the HERMES data
used here~\cite{DSEHS}. It is surprising to
find that it does not satisfy our conditions derived here. This unexpected finding
is probably due to the fact that DSEHS is
a NLO fitting result. A more careful investigation into this issue is desirable
but definitely beyond the scope of this article.\\

Since the value of $D^{\pi}_{Q}(Q^2)$ is a combination of the integrals of four FFs, it is instructive to
investigate the individual contribution of each FF. The results are listed in Table~\ref{Table:pion}. Among the
four FFs contributing to
$D^{\pi}_{Q}(Q^2)$, $D^{\pi^{+}}_{u}(Q^2)$, and $D^{\pi^{-}}_{d}(Q^2)$ are obviously the dominant ones, because both of them are favored FFs.
The FFs satisfying our condition Eq.~(\ref{constraint2}) all give quite large values of $D^{\pi^{+}}_{u}$ and $D^{\pi^{-}}_{d}$ ($\ge 0.5$).
On the contrary, the FFs which fail to meet the requirement of Eq.~(\ref{constraint2})
all result in smaller $D^{\pi^{+}}_{u}$ and $D^{\pi^{-}}_{d}$ ($<0.5$). In other words, our conditions seem to prefer the $u$
quark to be fragmented into $\pi^{+}$ rather than to other kinds of hadrons, such as $K^{+}$. \\

Note that the situation of $D^{\pi}_{S}(Q^2)$
is completely different. The NJL-Jet model, NL$\chi$QM, and AKK08 parametrization all lead to very small values of
$D^{\pi}_{S}(Q^2)$ when compared with other FFs. The NJL-Jet model and AKK08 parametrization both, however, violate our
conditions, but NL$\chi$QM meets the requirement of Eq.~(\ref{constraint2}). The reason is that the values for $D^{\pi}_{Q}(Q^2)$ of
NL$\chi$QM are substantially large than the other two. This implies that the value of $D^{\pi}_{Q}(Q^2)$ plays a more important
role to meet the criterion. We would like to emphasize that although both the NJL-Jet model and NL$\chi$QM are
chiral models with the same couple-channel jet algorithm, the results of the two models are significantly different.
In particular, while the NLJ-Jet model violates the constraints at all data points, NL$\chi$QM meets the requirement of
these constraints. In both models the corresponding model scales $Q_0^2$ are determined by fitting one of the FFs, usually, $D^{\pi^{+}}_{u}(z,Q^2)$.
One is able to obtain the FFs at arbitrary $Q^2$ by applying the QCD evolution. Here we use QCDNUM17~\cite{Botje:2010ay,DGLAP}.

\begin{table}[t]
\begin{tabular}{|c|c|c|c|c|c|c||c|c|c|}
\hline
Parametrization and model& $Q^2\,[\mathrm{GeV}^2]$ &$D_u^{\pi^{+}}$ & $D_d^{\pi^{+}}$ & $D_u^{\pi^{-}}$ & $D_d^{\pi^{-}}$& $D^{\pi}_{Q}$ &$D_s^{\pi^{+}}$ & $D_s^{\pi^{-}}$ & $D^{\pi}_{S}$ \\ \hline
HKNS(LO) &1.1931 & 0.80& 0.35 &0.35 &0.80& 5.76 & 0.35 &0.35 & 1.40 \\  \hline
HKNS(LO) &10.2355 & 0.62& 0.25 &0.25 &0.62& 4.35 & 0.25 &0.25 & 1.00 \\  \hline
\hline
HKNS(NLO)  &1.1931& 0.64 &0.21 &0.21&0.64&4.23&0.21&0.21&0.83\\ \hline
HKNS(NLO)&10.2355  & 0.52 &0.19 &0.19&0.52&3.53&0.19&0.19&0.74\\ \hline
\hline
DSS(LO) &1.1931 & 0.47 &0.20 & 0.20 & 0.53 & 3.38 & 0.17 & 0.17 & 0.67 \\  \hline
DSS(LO)&10.2355  & 0.44 &0.23 & 0.23 & 0.49 & 3.41 & 0.21 & 0.21 & 0.82 \\  \hline
\hline
DSS(NLO) &1.1931 & 0.45 &0.20 &0.20&0.51&3.31&0.17&0.17&0.67\\ \hline
DSS(NLO)&10.2355  & 0.41 &0.20 &0.20&0.46&3.12&0.18&0.18&0.74\\ \hline
\hline
NJL-Jet  &1.1931& 0.46 &0.12 &0.12&0.45&2.83&0.08&0.08&0.32\\ \hline
NJL-Jet &10.2355 & 0.38 &0.10 &0.10&0.38&2.38&0.07&0.07&0.28\\ \hline
\hline
NL$\chi$QM  &1.1931 & 0.68&0.22 &0.22&0.68&4.50&0.07&0.07&0.28\\ \hline
NL$\chi$QM &10.2355 & 0.54&0.16 &0.16&0.55&3.53&0.05&0.05&0.21\\ \hline
\hline
AKK08 &1.1931 & 0.26 &0.25 &0.25&0.26&2.51&0.02&0.02&0.07\\ \hline
AKK08 &10.2355 & 0.33 &0.31 &0.31&0.33&3.19&0.11&0.11&0.45\\ \hline
\hline
SKMA  &1.1931& 0.60 &0.24 &0.24&0.60&4.19&0.24&0.24&0.94\\ \hline
SKMA &10.2355 & 0.48 &0.18 &0.18&0.48&3.27&0.18&0.18&0.72\\ \hline
\hline
DSEHS  &1.1931&0.47 & 0.24 &0.24 &0.47&3.55&0.30 &0.30&1.21\\ \hline
DSEHS &10.2355 &0.39 & 0.21 &0.21 &0.39&2.97&0.24 &0.24&0.97 \\
\hline
\end{tabular}
\caption{The integrated FFs over $z$ for each channel at $Q^2=1.1931\,\mathrm{GeV}^2$ and $10.2355\,\mathrm{GeV}^2$. The integration range is from $z_\mathrm{min}=0.2$ to $z_\mathrm{max}$=0.8. }
\label{Table:pion}
\end{table}
\section{The constraint on the fragmentation functions of charged kaons}
In this section, we apply the same analysis to the kaon multiplicity. Again we rewrite Eq.~(\ref{Eq:kaon}) as follows:
\begin{equation}
S(x,Q^2)=\left[\frac{5M^{K}_{D}(x,Q^2)-D^{K}_{Q}(Q^2)}{D^{K}_{S}(Q^2)-2M^{K}_{D}(x,Q^2)}\right]Q(x,Q^2).
\label{Eq:S/Q_kaon}
\end{equation}
Furthermore, similar to the pion case, we require that the following relations to be held, considering the positiveness
of PDFs:
\begin{equation}
\frac{D^{K}_{S}(Q^2)}{2}<M^{K}_{D}(x,Q^2)\le\frac{D^{K}_{Q}(Q^2)}{5}\,\,\,\,\mathrm{or}\,\,\,\,
\frac{D^{K}_{Q}(Q^2)}{5}\le M^{K}_{D}(x,Q^2)<\frac{D^{K}_{S}(Q^2)}{2}.
\label{constraint:kaon}
\end{equation}

\begin{figure}[t]
\begin{tabular}{ccc}
\includegraphics[width=5.2cm]{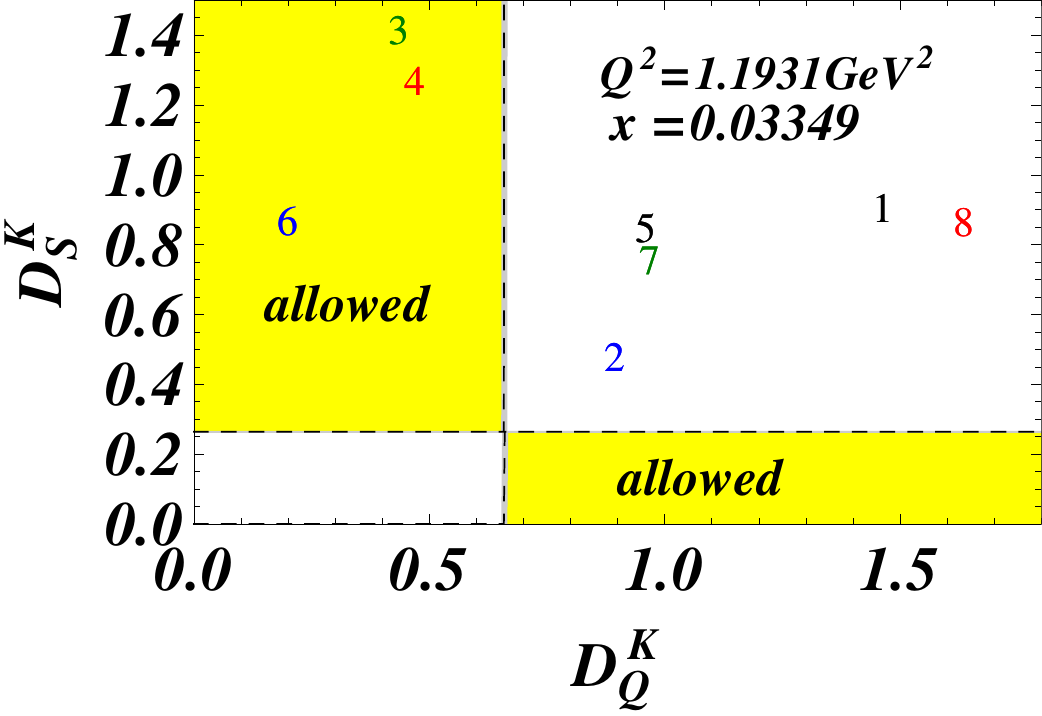}
\includegraphics[width=5.2cm]{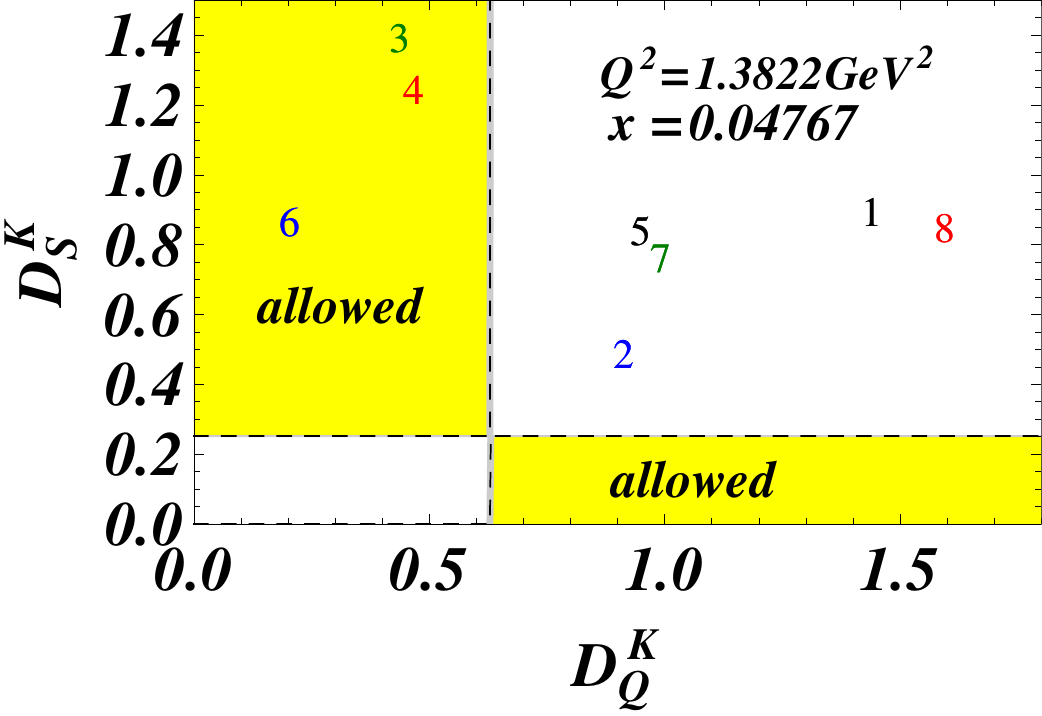}
\includegraphics[width=5.2cm]{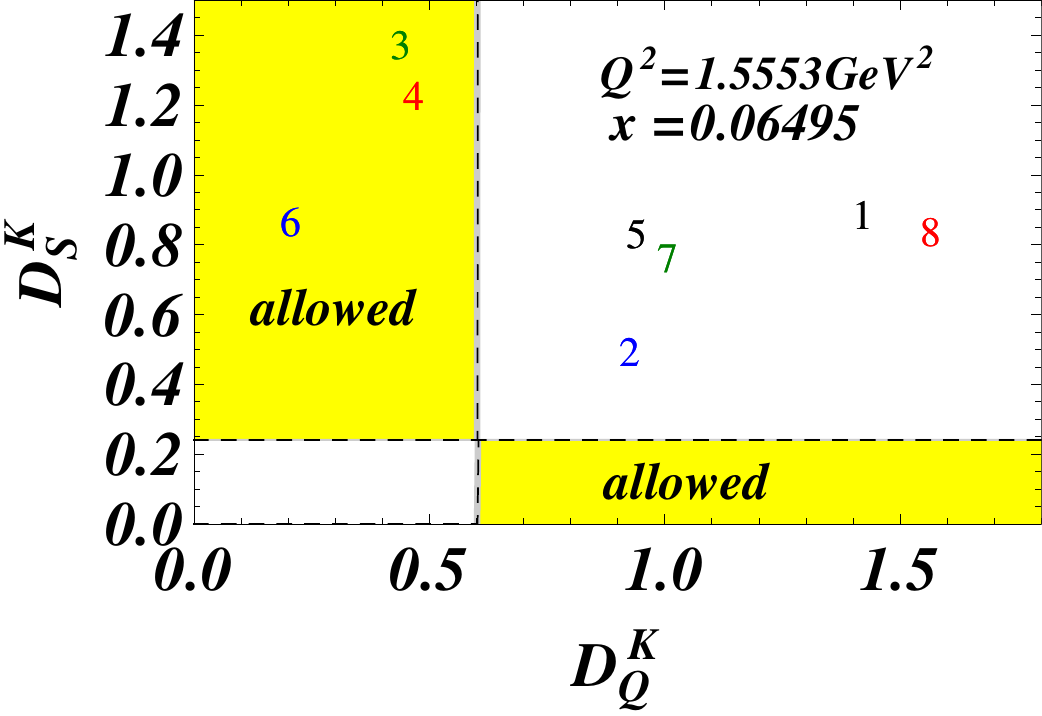}
\end{tabular}
\begin{tabular}{ccc}
\includegraphics[width=5.2cm]{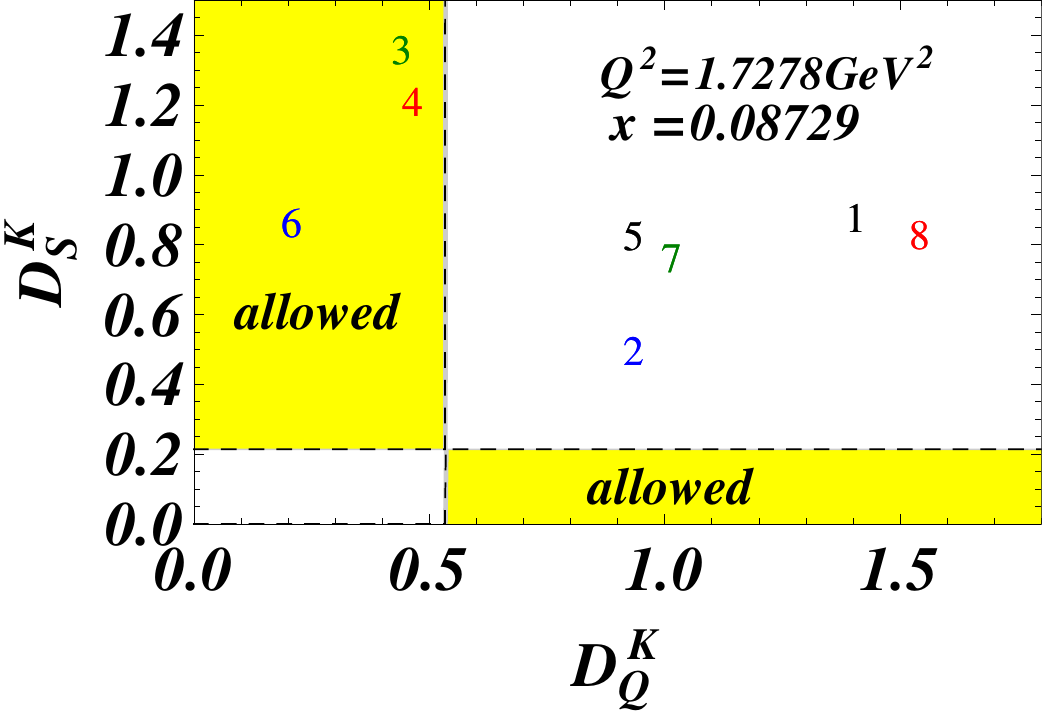}
\includegraphics[width=5.2cm]{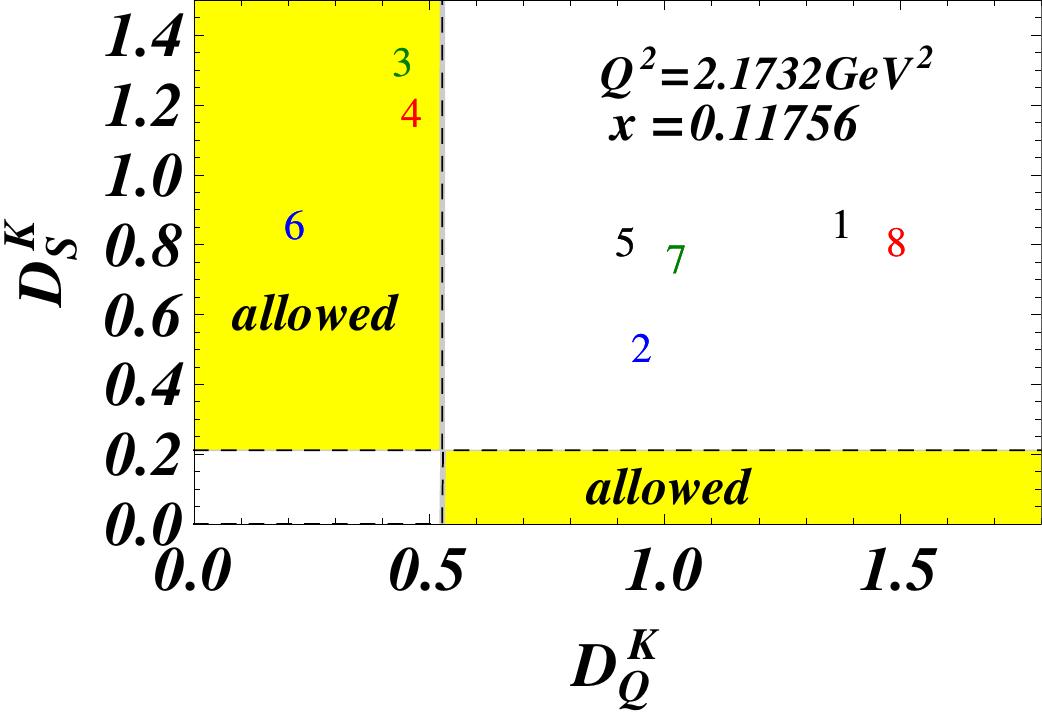}
\includegraphics[width=5.2cm]{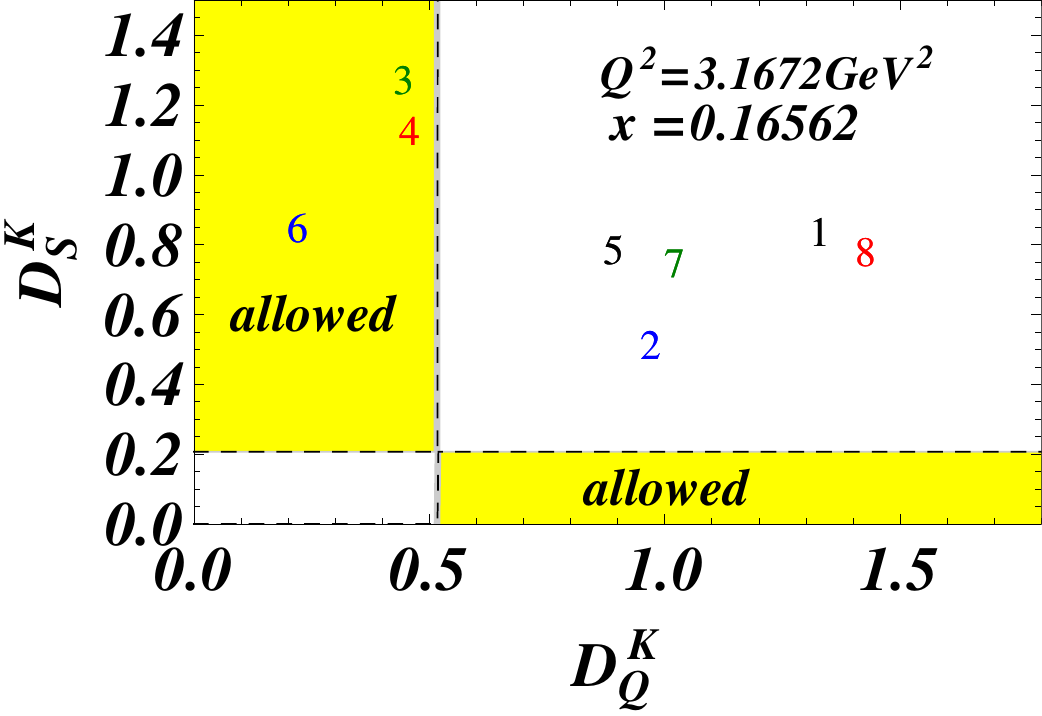}
\end{tabular}
\begin{tabular}{ccc}
\includegraphics[width=5.2cm]{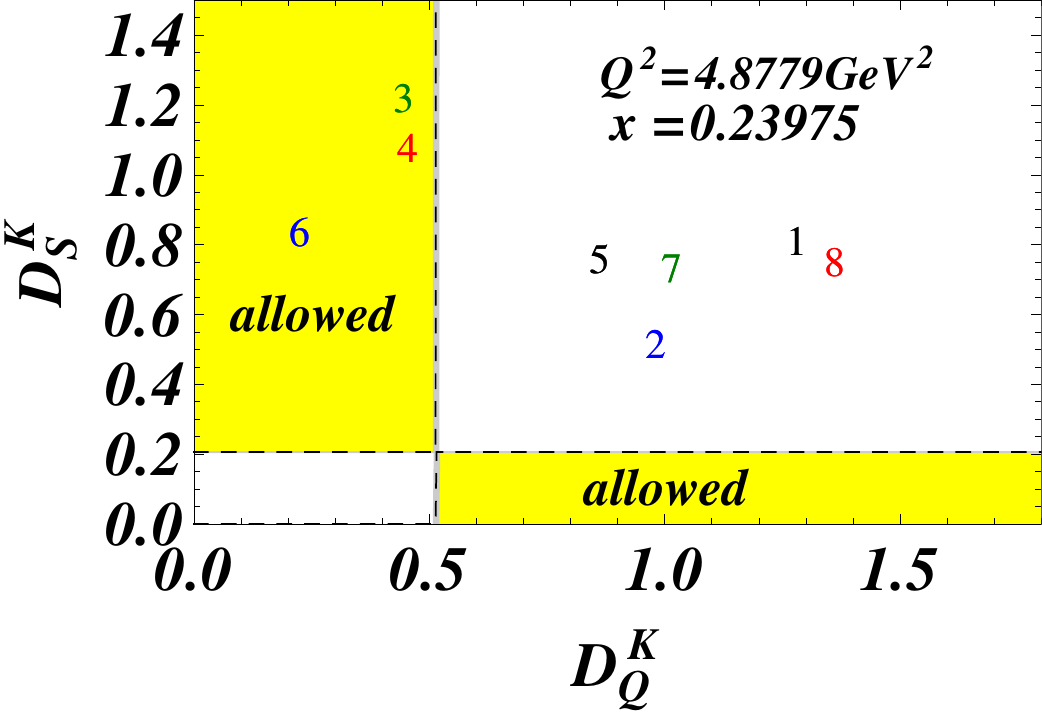}
\includegraphics[width=5.2cm]{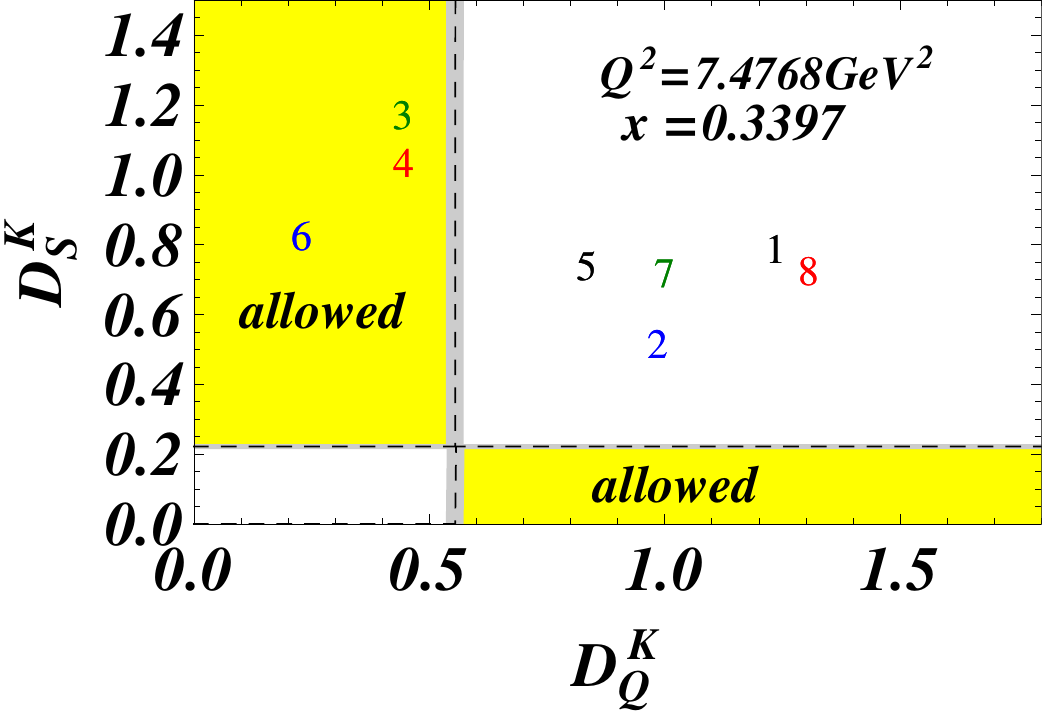}
\includegraphics[width=5.2cm]{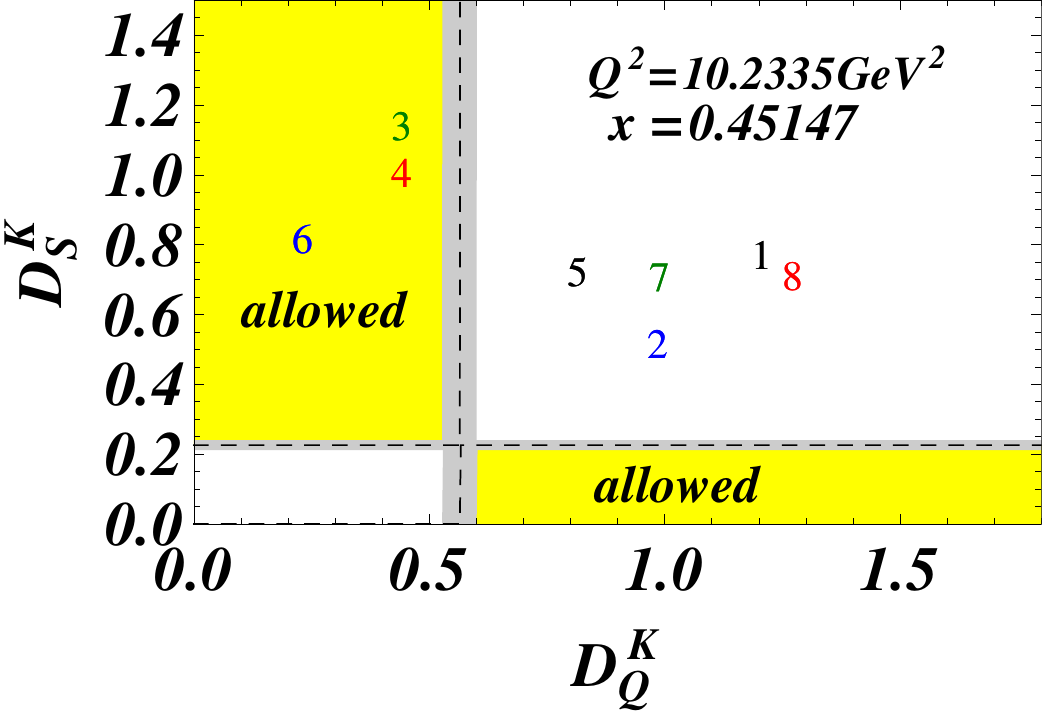}
\end{tabular}
\caption{
The values of $D^{K}_{Q}(Q^2)$ and $D^{K}_{S}(Q^2)$ defined in Eq.~(\ref{Eq:AB_K}) from various fragmentation functions: HKNS parametrization at LO (1), HKNS parametrization at NLO (2), DSS parametrization at LO (3), DSS parametrization at NLO (4),  NJL-Jet model (5), nonlocal chiral-quark model (6),  AKK08 parametrization (7), and SKMA parametrization (8). The yellow blocks represent the allowed regions experimentally. The grey bands stand for the areas corresponding to the estimated uncertainties of $M_{D}^{K}$.}
\label{Fig:kaon}
\end{figure}

We present the results of $D^{K}_{Q}(Q^2)$ and $D^{K}_{S}(Q^2)$ in Fig.~\ref{Fig:kaon}. The yellow block at r.h.s represents the region
satisfying $D^{K}_{Q}(Q^2)\ge 5M^{K}_{D}(x,Q^2)$ and $D^{K}_{S}(Q^2)<2M^{K}_{D}(x,Q^2)$. The l.h.s. block denotes the area for
$D^{K}_{Q}(Q^2)\le 5M^{K}_{d}(x,Q^2)$ and $D^{K}_{S}(Q^2)>2M^{K}_{d}(x,Q^2)$. The values of $D^{K}_{Q}$ and $D^{K}_{S}$ must be within either
block otherwise the value of $S(x,Q^2)/Q(x,Q^2)$ will turn to negative.
We notice that only the results of LO (3) and NLO (4) DSS parametrizations and NL$\chi$QM (6) satisfy the constraints given in
Eq.~(\ref{constraint:kaon}). They pass the test at all data points. In the contrast, other FFs fail to meet the criterion at
every data point. From Fig.~\ref{Fig:kaon}, we know the points corresponding to DSS parametrizations (3 and 4) are on
the right brink of the l.h.s. block but the NL$\chi$QM result (6) locates deep inside the r.h.s. block. It is obvious that the value of
$D^{K}_{Q}$ plays crucial role here. Unless its $D^{K}_{Q}(Q^2)$ is smaller than $0.5$, the FF would fail to make $S(x,Q^2)/Q(x,Q^2)$
positive. We find that $D^{K}_{Q}(Q^2)$ from DSS parametrizations are slightly below $0.5$, and $D^{K}_{Q}$ from NL$\chi$QM are even smaller,
around $0.2\sim0.3$. Although the value of $D^{K}_{S}$ plays minor role with respect to the criterion, but it will be vital in extracting
$S(x,Q^2)$ from the $M^{K}_{D}(x,Q^2)$ data. We notice that the DSS parametrizations produce relatively large $D^{K}_{S}(Q^2)$ ($\ge 1.0$).
The values of $D^{K}_{S}(Q^2)$ from the other FFs are all below $1.0$.

\begin{table}[t]
\begin{tabular}{|c|c|c|c|c|c|c||c|c|c|}
\hline
Model& $Q^2$ & $D_u^{K^{+}}$ & $D_d^{K^{+}}$ & $D_u^{K^{-}}$ & $D_d^{K^{-}}$& $D^{K}_{Q}$ &$D_s^{K^{+}}$ & $D_s^{K^{-}}$ &$D^{K}_{S}$ \\
\hline
HKNS(LO)&1.1931 & 0.18& 0.12 & 0.12 & 0.12 & 1.46 & 0.12 & 0.33 & 0.90 \\ \hline
HKNS(LO)& 10.2355 & 0.16& 0.10 & 0.10 & 0.10 & 1.21 & 0.10 & 0.29 & 0.77 \\ \hline
\hline
HKNS(NLO)&1.1931  & 0.14 &0.06 &0.06&0.06&0.89&0.06&0.18&0.48\\ \hline
HKNS(NLO)& 10.2355 & 0.14 &0.07 &0.07&0.07&0.98&0.07&0.18&0.51\\ \hline
\hline
DSS(LO)&1.1931  & 0.10 & 0.01 & 0.01 & 0.01 & 0.43 & 0.01 & 0.70 & 1.41 \\  \hline
DSS(LO)& 10.2355 & 0.09 & 0.02 & 0.02 & 0.02 & 0.44 & 0.02 & 0.55 & 1.13 \\  \hline
\hline
DSS(NLO)&1.1931  & 0.10 &0.01 &0.01&0.01&0.47&0.01&0.62&1.27\\ \hline
DSS(NLO)& 10.2355 & 0.09 &0.02 &0.02&0.02&0.44&0.02&0.49&1.00\\ \hline
\hline
NJL-JET&1.1931& 0.15 &0.05 &0.06&0.06&0.96&0.04&0.40&0.85\\ \hline
NJL-JET & 10.2355& 0.13 &0.05 &0.05&0.05&0.82&0.04&0.32&0.72\\ \hline
\hline
NL$\chi$QM &1.1931& 0.03 &0.02 &0.01&0.01&0.20&0.01&0.42&0.87\\ \hline
NL$\chi$QM & 10.2355& 0.03 &0.02 &0.02&0.02&0.33&0.01&0.39&0.81\\ \hline
\hline
AKK08 &1.1931& 0.20 &0.05 &0.02&0.05&0.97&0.10&0.28&0.76\\ \hline
AKK08 & 10.2355& 0.18 &0.05 &0.03&0.05&0.99&0.10&0.25&0.70\\ \hline
\hline
SKMA &1.1931& 0.13 &0.07 &0.07&0.07&0.93&0.07&0.27&0.67\\ \hline
SKMA & 10.2355 & 0.12 &0.06 &0.06&0.06&0.80&0.06&0.23&0.58\\
\hline
\end{tabular}
\label{Table:kaon}
\caption{The integrated FFs over $z$ for each channel at $Q^2=1.1931\,\mathrm{GeV}^2$ and $10.2355\,\mathrm{GeV}^2$. The integration range
is from $z_\mathrm{min}=0.2$ to $z_\mathrm{max}$=0.8. }
\end{table}

The individual contributions of each FF are listed in Table~IV. The favored ones, $u\to K^{+}$ and $s\to K^{-}$ are larger than the
unfavored ones as expected. We notice that, being compared with the other FFs, $D_{u}^{K^{+}}$ of NL$\chi$QM and DSS parametrizations, LO
and NLO, are substantially smaller. It is also worthy of mentioning that the $D_{s}^{K^{-}}(Q^2)$ of DSS parametrizations are particularly large.
Our observation is that the conditions derived here seem to prefer the FFs with large $D_{Q}^{\pi}(Q^2)$ but small $D_{Q}^{K}(Q^2)$.
In other words, our analysis shows that within the LO QCD calculations, the HERMES data suggests
that more $K^{-}$ meson to be fragmented from $s$ quark rather than $\bar{u}$ quark, and more $K^{+}$ to be fragmented from $\bar{s}$ quark
rather than $u$ quark. That is, most of the kaons should be fragmented from the quarks with strangeness. However we have to
emphasize that this observation is only qualitative and within the LO QCD analysis.

\section{Relations between the charged kaon and pion multiplicities}

In this section, we address the issue of consistency between the charged pion and kaon multiplicities.
From Eqs.~(\ref{Eq:S/Q_pion}) and (\ref{Eq:S/Q_kaon}), one obtains
\begin{equation}
\frac{S(x,Q^2)}{Q(x,Q^2)}=\frac{5M^{\pi}_{D}(x,Q^2)-D^{\pi}_{Q}(Q^2)}{D^{\pi}_{S}(Q^2)-2M^{\pi}_{D}(x,Q^2)}=
\frac{5M^{K}_{D}(x,Q^2)-D^{K}_{Q}(Q^2)}{D^{K}_{S}(Q^2)-2M^{K}_{D}(x,Q^2)}.
\label{Eq:SQ}
\end{equation}
Hence, one can determine $S(x,Q^2)/Q(x,Q^2)$ from $D^{\pi}_{Q}(Q^2)$, $D^{\pi}_{S}(Q^2)$, and $M^{\pi}_{D}(x)(x,Q^2)$. It is also
possible to employ $D^{K}_{Q}(Q^2)$, $D^{K}_{S}(Q^2)$, and $M^{K}_{D}(x,Q^2)$ to decide the values of $S(x,Q^2)/Q(x,Q^2)$. Naturally, the two
results should be consistent. Therefore, once the values of $D^{\pi,K}_{Q}(Q^2)$ and $D^{\pi,K}_{S}(Q^2)$ are known,
Eq.~(\ref{Eq:SQ}) becomes a relation between $M^{\pi}_{D}(x,Q^2)$ and $M^{K}_{D}(x,Q^2)$.

\begin{figure}
\begin{tabular}{c}
\includegraphics[width=7.2cm]{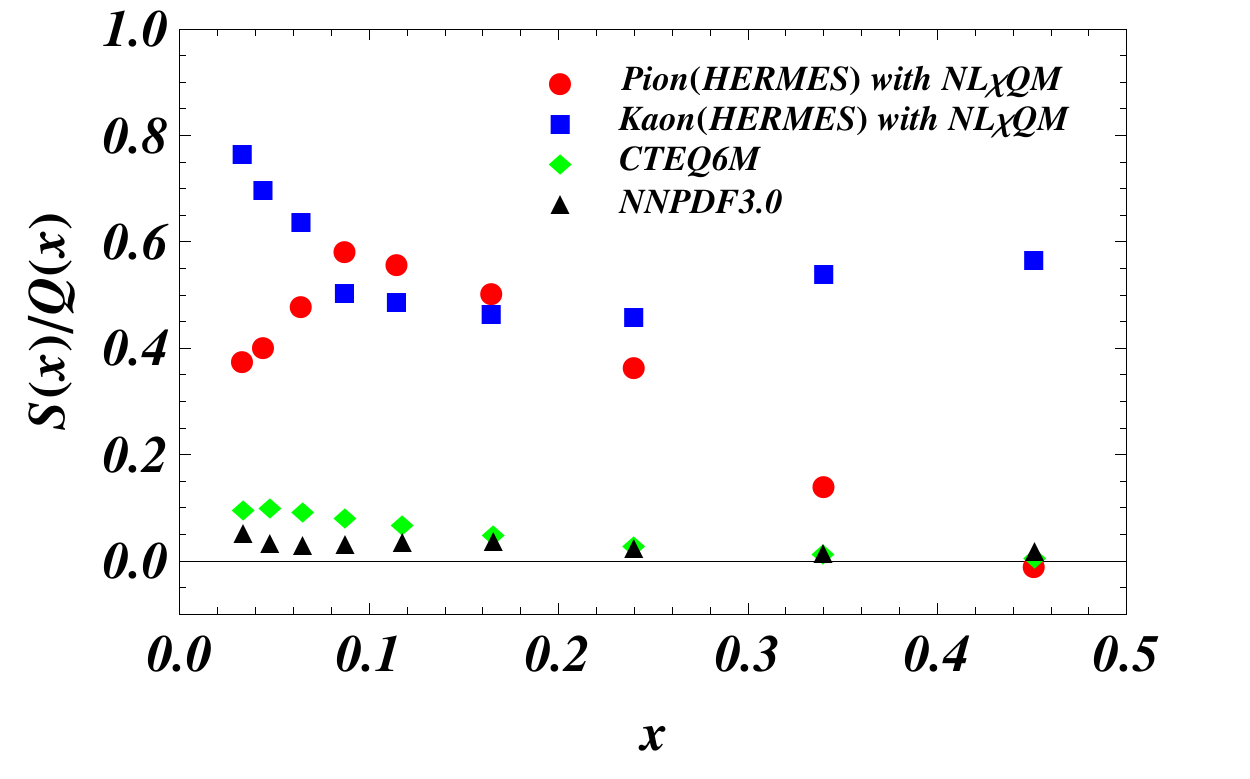}
\end{tabular}
\caption{$S(x,Q^2)/Q(x,Q^2)$ extracted from CTEQ6, NNPDF and the
results of $S(x,Q^2)/Q(x,Q^2)$ from HERMES data of $M^{\pi}_{D}$ and $M_{D}^{K}$ with the $D^{\pi,K}_{Q}$ and $D^{\pi,K}_{S}$
from the nonlocal chiral-quark model.}
\label{Fig:SQI}
\begin{tabular}{cc}
\includegraphics[width=7.2cm]{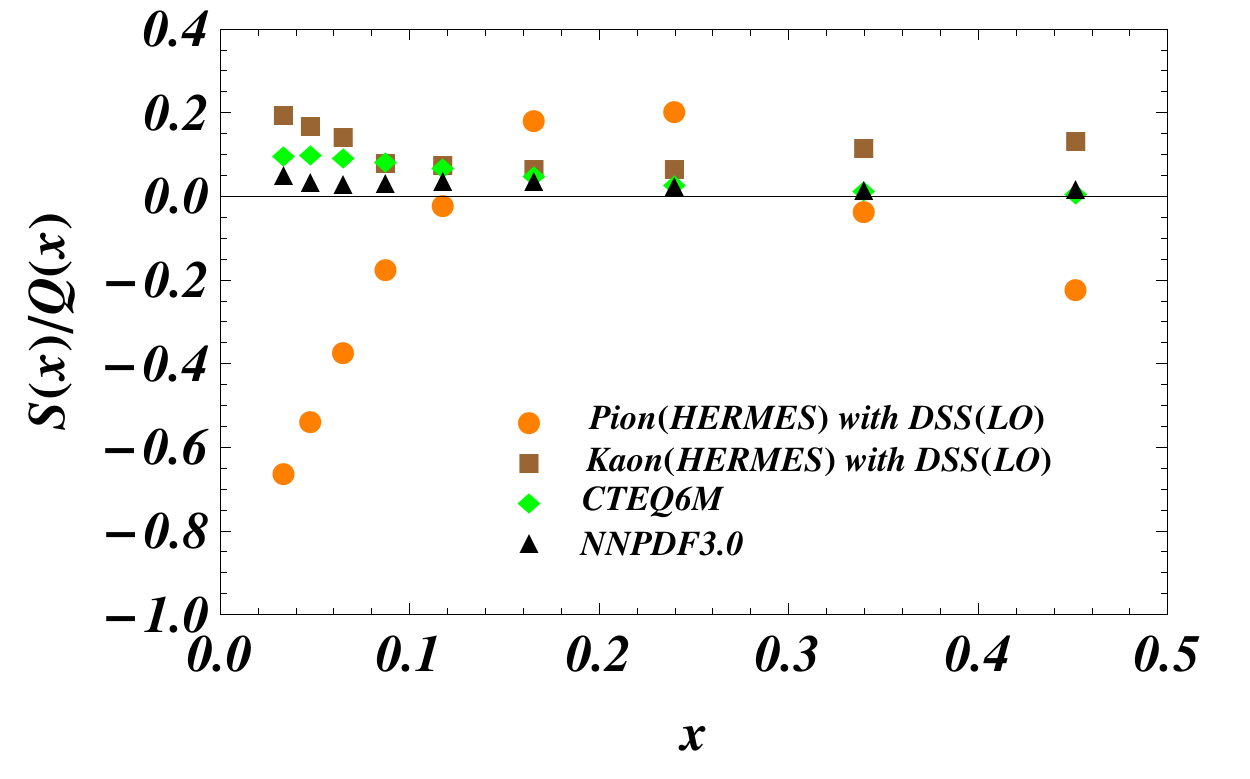}
\includegraphics[width=7.2cm]{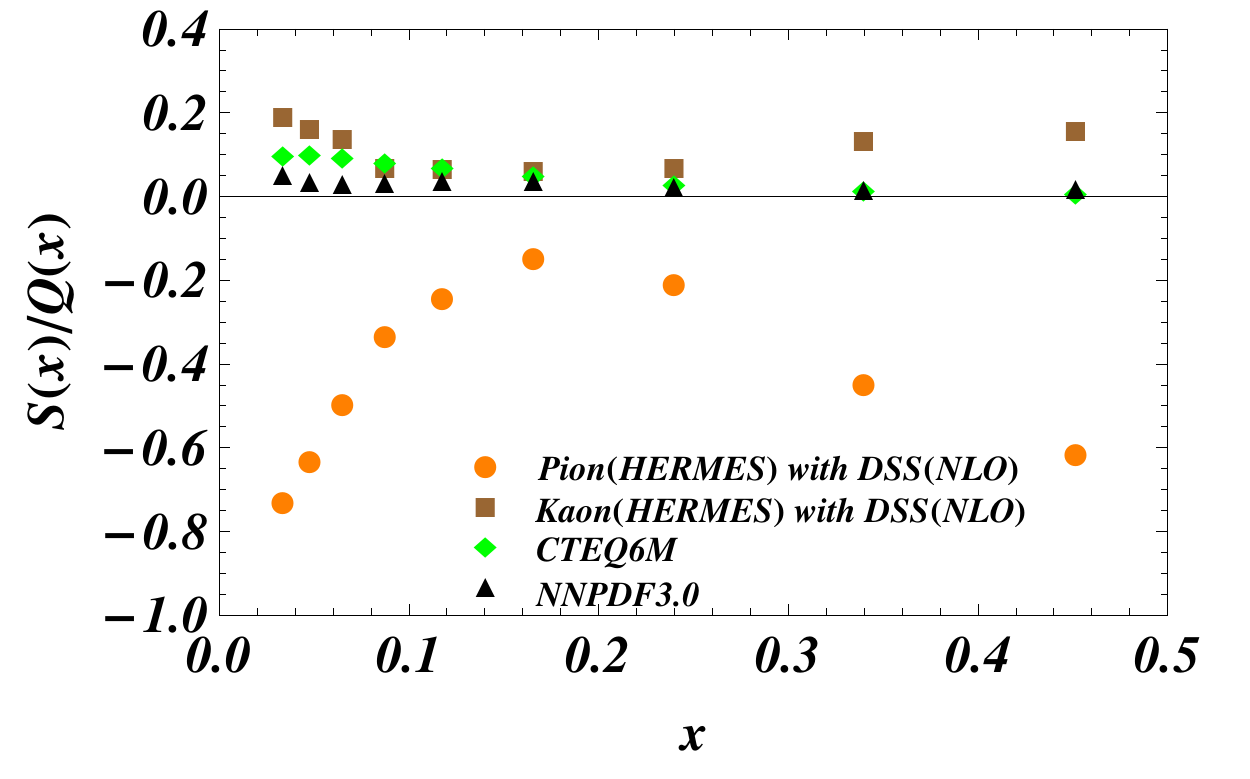}
\end{tabular}
\caption{$S(x,Q^2)/Q(x,Q^2)$ extracted from CTEQ6, NNPDF and the
results of $S(x,Q^2)/Q(x,Q^2)$ from HERMES data of $M^{\pi}_{D}$ and $M_{D}^{K}$ with the $D^{\pi,K}_{Q}$ and $D^{\pi,K}_{S}$ from DSS LO (left) and NLO (right) parametrizations.}
\label{Fig:SQII}
\begin{tabular}{cc}
\includegraphics[width=7.2cm]{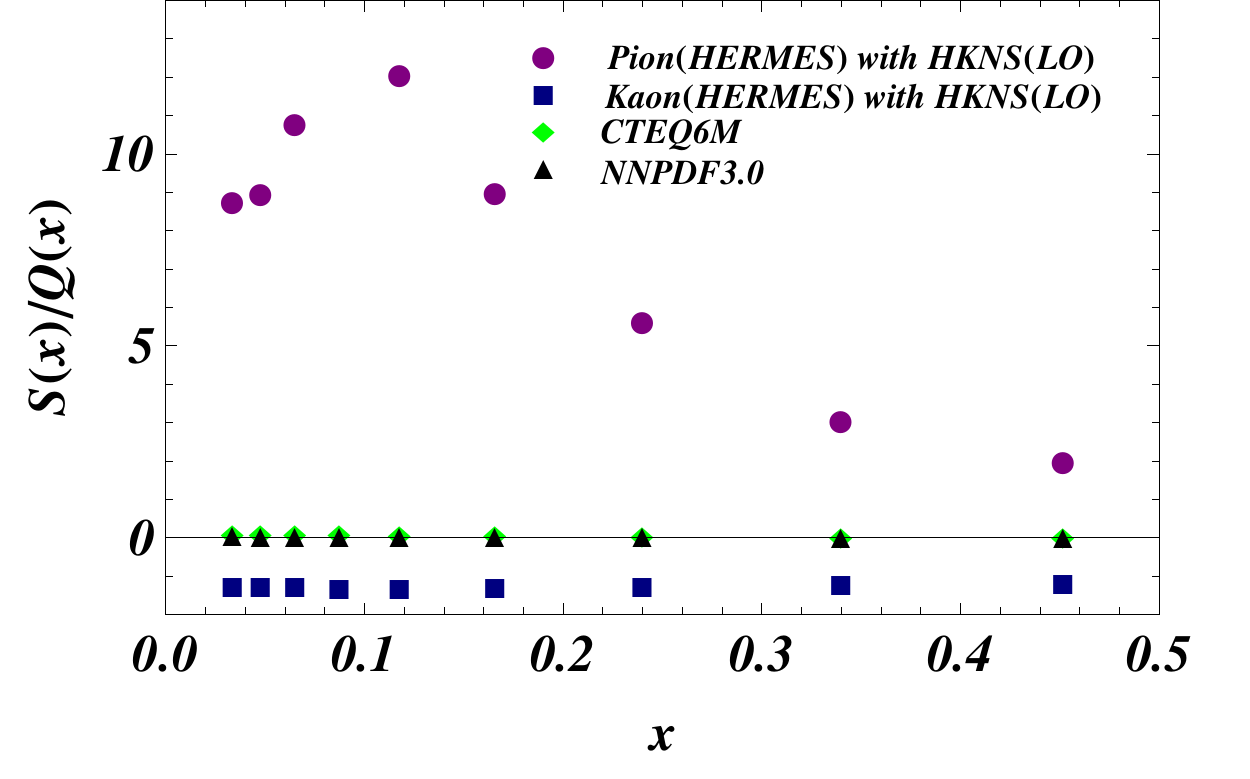}
\includegraphics[width=7.2cm]{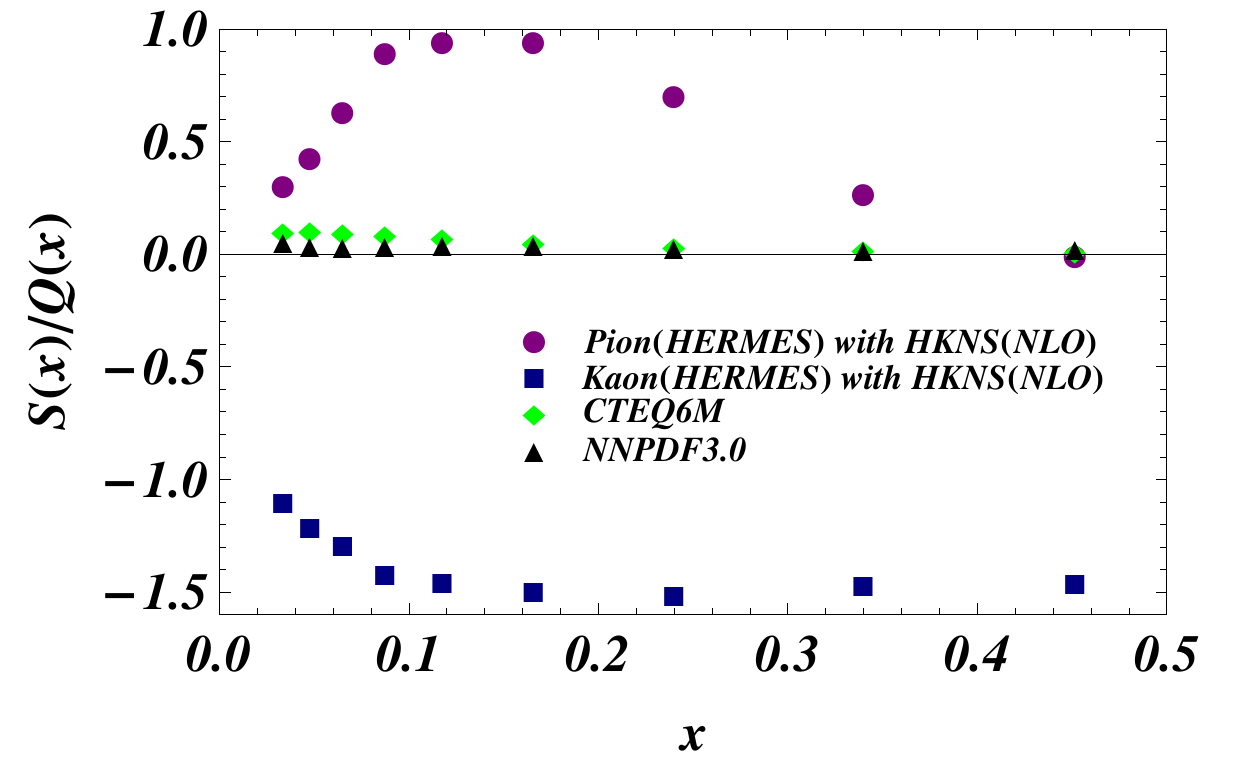}
\end{tabular}
\caption{$S(x,Q^2)/Q(x,Q^2)$ extracted from CTEQ6, NNPDF and the
results of $S(x,Q^2)/Q(x,Q^2)$ from HERMES data of $M^{\pi}_{D}$ and $M_{D}^{K}$ with the $D^{\pi,K}_{Q}$ and $D^{\pi,K}_{S}$
from HKNS LO (left) and NLO (right) parametrizations.}
\label{Fig:SQIII}
\end{figure}

Furthermore, one can directly obtain $S(x,Q^2)/Q(x,Q^2)$ from available PDFs.
Here we take two parametrizations as examples: CTEQ6M~\cite{CTEQLO} and NNPDF3.0~\cite{NNPDF}. One may expect naively that
$S(x,Q^2)/Q(x,Q^2)$ from these approaches should be all consistent with each other. However our analyses shows otherwise. As a matter of fact, from
Fig.~\ref{Fig:SQI}, one finds that the results of $S(x,Q^2)/Q(x,Q^2)$ derived from $M^{\pi}_{D}(x,Q^2)$ and
$M^{K}_{D}(x,Q^2)$ are different from each other. In reaching the results in Fig.~\ref{Fig:SQI}, the values of
$D^{\pi,K}_{Q}(Q^2)$ and $D^{\pi,K}_{S}(Q^2)$ produced from NL$\chi$QM are employed, because it is the only one which is able to make
$S(x,Q^2)/Q(x,Q^2)$ to be always positive among the chosen FFs in this study.\\

For the data taken at the low $x$ and $Q^2$ values, the $S(x,Q^2)/Q(x,Q^2)$ extracted from $M^{\pi}_{D}$ are smaller than the ones from
$M^{K}_{D}$. Interestingly, however, the situation is changed as $x$ and $Q^2$ increase.
The results from $M^{\pi}_{D}$ decrease, whereas the results from $M^{K}_{D}$ are very stable and are around $0.5$.
Only at the data points (D) and (E) ($0.3\le x \le0.5$), the two results are
consistent. They are far away from the results directly from the PDFs, either CTQE6M or NNPDF3.0. $S(x,Q^2)/Q(x,Q^2)$ from CTQE6M and
NNPDF3.0 are very close to each other, but they are much smaller than the ones from charged mesons multiplicities except at data
points (H) and (J), where the results from the pion multiplicities are quite close to the results from the PDFs. On the contrary,
the results from $M^{K}_{D}$ are always larger than the one from PDFs.

In Fig.~\ref{Fig:SQII}, we present similar analyses using $D^{\pi,K}_{Q}(Q^2)$ and $D^{\pi,K}_{S}(Q^2)$ from the LO and NLO DSS
parametrizations. It is found that $S(x,Q^2)/Q(x,Q^2)$ extracted form $M^{K}_{D}$ become quite close to the values
from CTEQ6M and NNPDF3.0. After a close look, one realizes that the values of $S(x,Q^2)/Q(x,Q^2)$ from
$M^{K}_{D}$ are still larger than the PDF results. Remember that $D^{K}_{S}$ of the DSS parametrizations, both at
LO and NLO, are significantly larger than others.
Since $M^{K}_{D}(x)$ is
related to $S(x)$ by the product $S(x,Q^2)D^{K}_{S}(Q^2)$, consequently, $S(x,Q^2)$ will increase if $D^{K}_{S}(Q^2)$
decreases, and vice versa. This explains why the $S(x,Q^2)/Q(x,Q^2)$ from $M^{K}_{D}$ with the NL$\chi$QM value for
$D^{K}_{S}$ is much larger than the ones
with the DSS values, because the NL$\chi$QM value of $D^{K}_{S}$ is only about $0.8$ but the DSS ones are around $1.2$ (LO) or $1.1$ (NLO).
Moreover, one needs even larger values for $D^{K}_{S}(Q^2)$ to reproduce the values of CTEQ6M and NNPDF3.0. On the other hands,
due to the fact that they are negative, i.e. physically unacceptable, one cannot extract $S(x,Q^2)/Q(x,Q^2)$ from
$M^{\pi}_{D}$ with the DSS values of $D^{\pi}_{Q,S}$.

Next, we turn to the case of the HKNS parametrizations. Let us first take a look of the left panel of
Fig.~\ref{Fig:SQIII}, in which $S(x,Q^2)/Q(x,Q^2)$ are extracted from $M^{\pi,K}_{D}$ with the LO HKNS values of
$D^{\pi}_{Q}(Q^2)$ and $D^{\pi}_{S}(Q^2)$ as inputs.
The results from $M^{\pi}_{D}(x,Q^2)$ are positive with enormous magnitude. On the other hand, the result remains positive but much
smaller, when NLO HKNS $D^{\pi}_{Q}(Q^2)$ and $D^{\pi}_{S}(Q^2)$ are used.
Unfortunately, both LO and NLO HKNS parametrizations cannot make $S(x,Q^2)/Q(x,Q^2)$ from $M^{K}_{D}$ be positive at all data points.

We conclude that, in the LO QCD analysis and with the $D^{\pi,K}_{Q,S}(Q^2)$ of the FFs chosen in our study,
there is inconsistency between $S(x,Q^2)/Q(x,Q^2)$ derived from the pion and kaon multiplicity data.
Moreover, even the same data set of $ M^{\pi, K}_{D}(x,Q^2)$ is used, the generated $S(x,Q^2)/Q(x,Q^2)$ from
various employed FFs are different from each other. It means that the extraction of the strange-quark PDF from the charged hadron
multiplicities of SIDIS depends strongly on the the choice of the FFs. Such uncertainty should be taken into account in the extraction
o $S(x,Q^2)$ as conducted in \cite{kaon2}.

\section{Charged meson multiplicities and parton distribution functions}
In previous section, we find that the values of $S(x,Q^2)/Q(x,Q^2)$ directly taken from CTEQ6M and NNPDF3.0
differ from the ones extracted from the HERMES data significantly.
Because of this discrepancy, naturally we would like to see with what values of $D^{\pi,K}_{Q}(Q^2)$ and $D^{\pi,K}_{Q}(Q^2)$
can one arrives at consistent results.
From Tables~\ref{Table:pion} and IV, we notice that the variations of $D^{\pi,K}_{Q}(Q^2)$ and $D^{\pi,K}_{S}(Q^2)$ with respect to the
change of $Q^2$ are very mild. With this observation as well as the assumption that $D^{\pi,K}_{Q}(Q^2)$ and
$D^{\pi,K}_{S}(Q^2)$ are constants, then we find
\begin{equation}
M^{\pi}_{D}(x,Q^2)\approx \frac{Q(x,Q^2)D_{Q}^{\pi}+S(x,Q^2)D_{S}^{\pi}}{5Q(x,Q^2)+2S(x,Q^2)},\,\,\,\,
M^{K}_{D}(x,Q^2)\approx \frac{Q(x,Q^2)D_{Q}^{K}+S(x,Q^2)D_{S}^{K}}{5Q(x,Q^2)+2S(x,Q^2)}.
\label{appro}
\end{equation}
Thus, one can fit $D^{\pi,K}_{Q}$ and $D^{\pi,K}_{S}$ from the data of $M^{\pi}_{D}(x,Q^2)$ and $M^{K}_{D}(x,Q^2)$ using the values
of $S(x,Q^2)$ and $Q(x,Q^2)$ taken from certain parametrizations of PDFs. The numbers in Tables.~\ref{Table:pion} and \ref{Table:kaon}
together with the results of the fit may shed some light in understanding the puzzle and explaining why the values of
$S(x,Q^2)/Q(x,Q^2)$ extracted from $M^{\pi}_{D}$ and $M^{K}_{D}$ in the previous section are so different from the ones directly
from the PDFs.\\

Let us focus the pion case first. It is a surprise to find that the fit values of $D^{\pi}_{S}$ are enormous in both
cases of CTEQ6M and NNPDF3.0 parametrizations. In other words, in order to explain the data of HERMES pion multiplicity with
CTEQ6M or NNPDF3.0, one needs to put $D^{\pi}_{S}\gg D^{\pi}_{Q}$. However, as we have mentioned already in Section II, $D^{\pi}_{Q}$
should be far more larger than $D^{\pi}_{S}$, since $D^{\pi}_{Q}$ contains the integrals of the favored fragmentation functions
$D_{u}^{\pi^{+}}(z,Q^2)$ and $D^{\pi^{-}}_{d}(z,Q^2)$. From Table.~\ref{Table:pion} one sees that $D^{\pi}_{Q}(Q^2)$ are indeed
much larger than $D^{\pi}_{S}(Q^2)$ in magnitudes. That is the reason why the derived $S(x,Q^2)/Q(x,Q^2)$ from
$M^{\pi}_{D}(x,Q^2)$ are so different
from the PDF values. More careful comparison shows that $D^{\pi}_{Q}$ in Table~\ref{Table:pion} are all larger than the fit
values of $D^{\pi}_{Q}$ shown in Table\ref{Table:PDF}. On the contrary, all $D^{\pi}_{S}$ in Table~\ref{Table:pion}
are all smaller than the fit values of $D^{\pi}_{S}$ by one order magnitude. Hence, one expects the values
of $S(x,Q^2)/Q(x,Q^2)$ extracted from the HERMES pion multiplicity with the values of $D^{\pi}_{Q}(Q^2)$ and $D^{\pi}_{S}(Q^2)$ from
Table~\ref{Table:pion} are much larger than those from CTEQ6M and NNPDF3.0. This has been verified in
Figs.~\ref{Fig:SQI}, \ref{Fig:SQII}, and \ref{Fig:SQIII}.

Next, for the kaon case, we find the fit values of $D^{K}_{S}$ are significantly larger than the values listed in
Table~\ref{Table:kaon}. The fit values of $D^{K}_{Q}$ are close to those produced from the DSS Parametrizations and
NL$\chi$QM. As a matter of fact, the DSS parametrizations generate the largest values of $D^{K}_{S}(Q^2)$ among our choices of the FFs,
but their values are still less than the half of the fit values shown in Table~\ref{Table:PDF}. Since $M^{K}_{D}$ is related
to $D^{K}_{S}$ by the product of $D^{K}_{S}$ and $S(x,Q^2)$, therefore, $S(x,Q^2)$ extracted from HERMES kaon data are always too large
compared with those from CTEQ6M and NNPDF3.0. The values of $D^{K}_{S}$ from NL$\chi$QM are smaller than the
DSS ones, consequently the resultant $S(x,Q^2)/Q(x,Q^2)$ are larger
than that
of DSS. This fact is transparent by taking a closer look of Figs.~\ref{Fig:SQI} and \ref{Fig:SQII}.

We also try to carry out the fit using only the data points with $Q^2\ge $ 2 GeV$^2$. The results of $D_{Q}^{\pi}$ and $D_{Q}^{K}$ does not
change much. On the other hand, the values of $D_{S}^{\pi}$ become much smaller.
However, even when only the data points of $Q^2\ge $ 2 GeV$^2$
are used, $D_{S}^{\pi}$ is still larger than $D_{Q}^{\pi}$. Furthermore we find $D_{S}^{K}$ also becomes smaller.
In particular, in the case of NNPDF3.0 (LO) $D_{S}^{K}$ even turns out to be negative!

Our result suggests that HERMES pion data are unlikely to be compatible with CTEQ6M and NNPDF3.0 parametrizations
within the framework of LO QCD analysis. It is anticipated that this scenario persists for other FFs than those used
in our study. This is because the favored FFs are definitely larger than the unfavored ones, hence no FF would give
$D^{\pi}_{S}\gg D^{\pi}_{Q}$. It is interesting to see whether the situation will be changed if other PDFs are adopted.
We also find that to reproduce the HERMES
data of $M^{K}_{D}$ with CTEQ6M or NNPDF3.0, one needs large $D^{K}_{S}$ and small $D^{K}_{Q}$.
Furthermore, none of our
chosen FFs would really match this criteria although the DSS parametrizations seems to be the most promising ones since
they are at the edge of fulfilling this goal.

\begin{table}[t]
\begin{tabular}{|c|c|c|c|c|}
\hline
FF& Data &CTEQ6M~\cite{CTEQLO}& NNPDF3.0 (LO)~\cite{NNPDF}& NNPDF3.0 (NLO)~\cite{NNPDF} \\ \hline
$D^{\pi}_{Q}$ &A-J &2.719 &3.305 &2.343  \\ \hline
$D^{\pi}_{S}$ &A-J&13.655&14.521&24.352\\ \hline
$D^{K}_{Q}$ &A-J&0.330 &0.483 &0.256  \\ \hline
$D^{K}_{S}$ &A-J&3.655&3.547&5.786\\ \hline
$D^{\pi}_{Q}$ &E-J &3.293 &3.337 &3.270  \\ \hline
$D^{\pi}_{S}$ &E-J&4.304&4.832&5.126\\ \hline
$D^{K}_{Q}$ &E-J&0.518 &0.537 &0.516  \\ \hline
$D^{K}_{S}$ &E-J&0.311&-0.139&0.381\\
\hline
\end{tabular}
\caption{The values of $D^{\pi}_{Q}$, $D^{\pi}_{S}$ , $D^{K}_{Q}$, and $D^{K}_{S}$, fitted from HERMES charged pion multiplicities with certain PDFs as inputs.}
\label{Table:PDF}
\end{table}

\section{Summary}
In summary, we used the HERMES SIDIS data of $M^{\pi}_{D}$ and $M^{K}_{D}$ to derive conditions
on the FFs and find that only NL$\chi$QM satisfy those conditions among all publicly available FFs. The preferred regions
of favoured $D_{Q}^{\pi,K}$ and $D_{S}^{\pi,K}$ should meet the inequalities of
$D^{\pi}_{Q}\gg D_{Q}^{K}$ and $D_{S}^{K}\gg D_{Q}^{K}$. This is consistent with the naive expectation from the
suppression of non-strange quark ($u$) fragmentation into $K^{+}$ (u$\bar{s})$ because of the production of $\bar{s}s$ pair.
Furthermore, we have shown that there exists inconsistency between the results of $S(x,Q^2)/Q(x,Q^2)$ extracted from the HERMES data of
charged pion and kaon multiplicities, if we use the FFs of NL$\chi$QM. We also find that the HERMES pion data is unlikely to be
compatible with the CTEQ6M and NNPDF3.0 as the PDFs even without referring any specific FFs. Our current study in this article is based
on two assumptions: 1) the leading order QCD formula for the multiplicity and  2) isospin symmetric nucleon
PDFs . To go beyond these two assumptions, it is necessary to extend our analysis substantially and will be reported elsewhere.

\section*{Acknowledgments}
We would like to thank Gunar Schnell for helpful comments and suggestions.
C.~W.~K.~ and D.~J.~Y.~ are supported by the grant 102-2112-M-033-005-MY3 from Ministry of Science and Technology (MOST) of Taiwan. F.~J.~J.~ is supported by MOST of Taiwan (grant No. 102-2112-M-003-004-MY3).


\end{document}